\documentstyle[referee]{l-aa}

\begin{document}

   \thesaurus{04(03.20.4;04.19.1;08.04.1;09.04.1;10.06.2;10.19.3)}

   \title{A new optical reddening model for the Solar neighborhood}

   \subtitle{Galactic Structure through low-latitude starcounts from
   the Guide Star Catalogue}

   \author{R. A. M\'endez\inst{1} \& W. F. van Altena\inst{2}}

   \offprints{R. A. M\'endez}

   \institute{(1) European Southern Observatory,
   Karl-Schwarzschild-Stra$\beta$e 2, D-85748, Garching b. M\"unchen,
   Germany, Electronic mail: rmendez@eso.org \\
   (2) Yale University
   Observatory, P.O. Box 208101, New Haven, CT 06520-8101, USA,
   Electronic mail: vanalten@astro.yale.edu}

   \date{Received; accepted}

   \maketitle

\begin{abstract}
A new optical reddening model for the Solar neighborhood is
presented. The model makes use of the large-scale properties of the
dust layer in the Galaxy, and of the observed clumpiness in its
distribution, and it is used to compute starcounts {\it in} the plane
of the Galaxy, which are compared to the observed counts from the
Guide Star Catalogue (GSC) for a few representative regions having a
range of reddening values. These comparisons provide a good test case
in a wide variety of conditions with regards to both the amount and
distribution of the reddening material along the line-of-sight.

This is the first systematic study of low-Galactic latitude starcounts
from the perspective of a Galactic structure and reddening model, and
it is demonstrated that the model can be used to obtain meaningful
starcount estimates in the plane of the Milky-Way.

We find that it is possible to predict starcounts in the range $8 \le
V< 13.7$ with a mean accuracy of 15\% or better for total samples of
several thousand stars at $|b| < 10^{\degr}$, and for total reddening
at 1~kpc of up to $E(B-V) \sim 0.7$~mag.

Self-consistency checks indicate that a differential optical
absorption of 0.5~mag/kpc is adequate to reproduce, with our model,
the available reddening maps in the distance range $2 \le r \le
6$~kpc.

Our best-fit model to three (out of six) selected regions yields a
distance of the Sun from the symmetry plane of the Galaxy of
$Z_{\sun}=27 \pm 3 \, (3 \sigma)$~pc, while the disk's scale-length is
found to be $H_{\rm Disk}= 6 \pm 2 \, (3\sigma)$~kpc. While the
derived value for $Z_{\sun}$ is quite sensitive to the adopted
scale-height of disk stars, no big dependency of $H_{\rm Disk}$ on
this parameter is found. These values should be taken, of course, with
caution because of the inherent uncertainties in modeling optical
starcounts in the presence of patchy reddening material.

The accuracy of the reddening and starcounts model opens the
possibility of using it to estimate and compare observations of diffuse
starlight to constrain the major structural parameters of the Galaxy,
and to compute the distance distribution of stars self-consistently
with the starcounts, in order to derive corrections to absolute
parallax.

\keywords{Techniques: photometric -- Surveys -- Stars: distances -- {\it
ISM:} dust, extinction -- Galaxy: fundamental parameters -- Galaxy:
structure}
\end{abstract}

\section{Introduction}

The method of starcounts as applied to the study of the general
properties of the Milky Way has become a very effective way of
constraining global structural parameters for the Galaxy (Reid \&
Majewski 1993), and novel multivariate data analysis methods are being
applied to recent surveys to characterize the individual stellar
distributions of each population, which would be otherwise
unrecognizable (Chen 1996, 1997).

Evidently, most previous comparisons between observed \& predicted
starcounts have avoided the large-extinction, low-Galactic latitude
regions (Bahcall 1986), where the dominant factor determining the
observed counts is the distribution of interstellar reddening, whose
complex nature would make such an analysis very difficult.

However, there are a number of situations where one may want to
evaluate the expected number of stars at low-Galactic latitudes. A few
examples are the need for estimating diffuse starlight in the plane of
the Milky-Way for comparisons with observations (Mattila 1980a,b, van
der Kruit 1986, Toller 1990), the evaluation of the corrections to
absolute proper-motion and parallax (van Altena 1974, Monet 1988), and
the evaluation of the expected kinematic distribution of stars at
low-Galactic latitude for comparisons with the upcoming large
digitized surveys (Humphreys 1993).

M\'endez (1995, also M\'endez \& van Altena 1996) has developed a
Galactic Structure model, capable of predicting starcounts {\it
self-consistently} with the Kinematic distribution of stars for any
line-of-sight. Evidently, in order to evaluate star --and color--
counts, as well as the expected kinematic properties of stars at
low-Galactic latitudes, we need to provide such a model with the
proper reddening values. For certain applications, one could use
specifically pre-determined values of reddening as a function of
Heliocentric distance (i.e., a {\it reddening map}). However, in many
instances, evaluations need to be performed at locations where no
specific determination of the reddening has been made, and in such
cases, we need to resort to the available low-latitude reddening
surveys as a means of estimating the reddening for that particular
location.

For the purposes outlined above, we have developed a reddening model,
based on what we know about the clumpy nature of the reddening
material in the Solar neighborhood from different surveys. We have
incorporated also what it is known about its large-scale structure
from those surveys, in order to interpolate the reddening values at
locations different from those studied in the available literature. In
Sect.~2 we introduce the reddening model in detail, Sect.~3 describes
our comparisons to starcounts at low-Galactic latitudes from the Guide
Star Catalogue, and in Sect.~4 we present our main conclusions.

\section{The reddening model}

\subsection{Introduction}

While interstellar absorption, in general, is caused by the mere
appearance of material along the line-of-sight (diffuse gas and dust
grains), the observed {\it selective} absorption (i.e., the variation of
absorption as a function of wavelength) is caused by {\it dust
particles} that have sizes on the order of the (optical) wavelength of
light. 

The flux, $F_{\lambda}(\vec{r})$ measured by an observer at a distance
$|\vec{r}|$ of a source located at position $\vec{r}$, and whose flux
at the Earth, in the absence of extinction, is $F_{\lambda}^{\rm o}$,
is given by $F_{\lambda}(\vec{r}) = F_{\lambda}^{\rm o} \, {\rm exp
}\left( -\tau_{\lambda}(\vec{r}) \right)$, where
$\tau_{\lambda}(\vec{r})$ is the (dimensionless) optical-depth of the
source along the path (Spitzer 1978). The optical-depth is usually
expressed in terms of the number volume density of the absorbing
material $n(\vec{r}) \mbox{ particles/cm$^3$}$, and the average
extinction cross-section per particle, $\kappa_{\lambda} \mbox{
cm$^2$/particle}$, (Mihalas \& Binney 1981) via the equation:

\begin{equation}
\tau_{\lambda}(\vec{r}) = \kappa_{\lambda} \, \int_0^{|\vec{r}|}
n(\vec{r'}) \, {\rm d}|\vec{r'}|
\end{equation}

Now, the absorption (in magnitudes) suffered by the light from this
source is given by $A_{\lambda}(\vec{r}) = -2.5 \, log
\left( F_{\lambda}(\vec{r}) / F_{\lambda}^{\rm o} \right) \approx 1.086 \,
\tau_{\lambda}(\vec{r})$.
 
The reddening to position $\vec{r}$, $E(ci)(\vec{r})$, is simply
related to $A_{\lambda}$ through the ratio of total-to-selective
absorption, defined as $R_{\lambda} = A_{\lambda}(\vec{r}) / \left(
A_{\lambda}(\vec{r}) - A_{\lambda r}(\vec{r}) \right) =
A_{\lambda}(\vec{r}) / E(ci)(\vec{r}) = \kappa_{\lambda} / \left(
\kappa_{\lambda} - \kappa_{\lambda 1} \right)$, where $ci$ refers to a
specific color index, and $\lambda r$ refers to a fiducial wavelength
(Spitzer 1978). As is evident from the above expression, $R_{\lambda}$
is determined {\it only} by the wavelength dependence of the
extinction coefficient, and it is independent of position.\footnote{Of
course, this is valid only for monochromatic radiation, for
finite-width passbands $R_{\lambda}$ becomes a weak function of
spectral type (Guti\'errez-Moreno \& Moreno 1975), see Sect. 3.2,
Eq. (16)} Since most applications have involved the Johnson \&
Morgan's (1953) B and V passbands, it is customary to define
$R_{\lambda}$ in terms of the reddening $E(B-V)$, defined as
$(B-V)-(B-V)_{\rm o}$, where $(B-V)$ is the observed (reddened) color
of a star with intrinsic color $({\rm B-V})_{\rm o}$. Of course, any
other passbands could be implemented this way (provided that we know
the extinction curve). We have adopted a value of $R_{\rm v}$ equal to
3.2 (Mihalas \& Binney 1981, Scheffler 1982, Scheffler \& Els\"asser
1987, although see Sect. 3.2).

\subsection{The large-scale distribution of reddening material}

Parenago (1945) was the first to propose a functional variation for
the reddening in the Galactic plane; he showed that an
exponentially-decaying function away from the Galactic plane provided
an accurate description for the distribution of reddening material as
derived from the (photoelectric) color excesses of around 3\,000
stars. His suggestion has been validated by later studies, most
notably those of Allen (1954) on the statistical distribution of
different types of stars (mainly Wolf-Rayet, O and A stars, and White
Dwarfs), and by Ochsenbein's (1983) study of the Z-distribution of Am
stars from the Michigan Spectral Catalogue. More recent studies have
found that the reddening per unit of distance (which is proportional
to $d\tau_{\lambda}(\vec{r})/{\rm d}|\vec{r}|$), as derived from a
complete sample of open clusters closer than 2~kpc follows an
exponential decay as a function of height above the plane (Pandey \&
Mahra 1987), in agreement with the assumption that $n(\vec{r})$ itself
follows a vertical exponential distribution (Eq. (1)), and in
agreement with earlier results by Lyng{\aa} (1982) from a more
restricted sample of open clusters.

Following Parenago, we assume that the (large-scale) volume (particle)
density of the absorbing material, $n(\vec{r})$, follows a decaying
exponential away from the Galactic plane, as given by:

\begin{eqnarray}
n(\vec{r}) & = & n_{\rm o} \, {\rm exp }\left( -|Z|/h_{\rm red} \right) \\
Z & = & Z_{\sun} + |\vec{r}| \, \sin b 
\end{eqnarray}

where $h_{\rm red}$ is the scale-height of the reddening material defining
the large-scale distribution of reddening material in a slab of
exponential thickness $2 \times h_{\rm red}$, $Z$ is the distance from the
Galactic plane of an object with Heliocentric distance $|\vec{r}|$ and
Galactic latitude $b$, and $Z_{\sun}$ is the distance of the Sun from
the Galactic plane.

There have been several determinations for the scale-height of the
reddening material. In Parenago's (1945) pioneer work, he found a
practically constant value for $h_{\rm red}$ equal to $100 \pm 4
\mbox{ pc}$. More recently, Lyng{\aa} (1982) found equally probable
the values of $h_{\rm red}= 130$ and $h_{\rm red}= 270 \mbox{ pc}$,
although the results of Burstein \& Heiles (1982) toward the Galactic
Poles (which imply $A_{\rm v}< 0.1 \mbox{ mag}$) would favor the
smaller value with Lyng\aa's adopted value of $0.75 \mbox{ mag/kpc}$
for $A_{\rm v}$. Pandey \& Mahra (1987) found $h_{\rm red}= 160 \pm 20
\mbox{ pc}$ from their complete sample of nearby open
clusters. Additional evidence comes from direct measurements of the
thickness of both neutral Hydrogen (HI) and Carbon Monoxide (CO) in
the Galactic disk. It is known that $E(B-V)$ is approximately
proportional to the neutral Hydrogen column density (Heiles 1976
[particularly his Fig. 13a, and references therein], Burstein \&
Heiles 1978 [particularly their Fig. 1a]), therefore the thickness of
the reddening material should be approximately well sampled by the
distribution of HI itself. Different studies (Scheffler \& Els\"asser
1987) have obtained thicknesses of about 130 to 150 pc from
measurements of cold diffuse HI. On the other hand, Spitzer (1978)
estimates that $h_{\rm red} \approx 100 \mbox{ pc}$ from the
statistical properties of the dust layer within 1 kpc of the
Sun. Finally, since the correlation $E(B-V)$ {\it vs.} $N({\rm HI})$
involves a gas-to-dust ratio, it would seem reasonable to expect that
$E(B-V)$ is related also to the distribution of molecular
material. From a compilation of different CO surveys (Solomon et
al. 1979, Table 1), we find $h_{\rm red}({\rm CO}) \sim 40 \; to \; 70
\mbox{ pc}$, while from their own CO survey of giant molecular clouds
they find $h_{\rm red}({\rm CO}) \approx 65 \mbox{ pc}$. As a
compromise value between all these different determinations, we have
assumed in our reddening model a scale-height of 110~pc.

\subsection{Reddening at intermediate and high Galactic latitudes}

For intermediate-to-high Galactic latitudes ($|b| \ge 10^{\degr}$),
reddening through ``infinity'' has been estimated by using Burstein \&
Heiles (1982, BH thereafter) reddening maps (available in
computer-readable form). These reddening estimates are used in our
model to compute the reddening via the equation:

\begin{eqnarray}
\frac {E(B-V)(r,l,b)}{E(B-V)(l,b)_{\infty}} = & \nonumber \\
\left\{ \begin{array}{ll} 
1-{\rm exp }\left( {-\frac{r\, \sin b}{h_{\rm red}}} \right) \; \mbox {if $b > +10^{\degr}$} \\
\frac {1-{\rm exp }\left( -\frac{r\, \sin b}{h_{\rm red}} \right) }
{1 - 2 \, {\rm exp }\left( \frac{Z_{\sun}}{h_{\rm red}} \right) } \;
\mbox{if $b < -10^{\degr}$ and $r \, |\sin b| \le Z_{\sun}$} \\
\frac {1- 2 \, {\rm exp }\left( \frac{Z_{\sun}}{h_{\rm red}} \right) +
{\rm exp }\left( \frac{2 \, Z_{\sun} + r \, \sin b}{h_{\rm red}} \right) }
{1 - 2 \, {\rm exp }\left( \frac{Z_{\sun}}{h_{\rm red}} \right) } \;
\mbox{if $b < -10^{\degr}$ and $r \, |\sin b| > Z_{\sun}$}
\end{array}
\right.
\end{eqnarray}

where r is the Heliocentric distance for a point with Galactic
coordinates (l,b), and $E(B-V)(l,b)_{\infty}$ is the reddening at
``$r=\infty$'' from the BH maps. Equation (4) has been derived by
performing the line-of-sight integration to Heliocentric distance r on
our assumed density law for the absorbing material (Eq. (2) and (3)),
as required by Eq. (1)). It should be noted that, if the Sun is
located in the plane of symmetry of the Galaxy, then Eq. (4) reduces
to the simpler case $E(B-V)(r,l,b)= E(B-V)(l, b)_{\infty} \times (1 -
{\rm exp} \left( -r \, |\sin b|/h_{\rm red}) \right)$, valid for any
value of the Galactic latitude (as long as $|b| \ge 10^{\degr}$).

Since the BH maps do not cover the entire sky, for regions where there
is no data available, we have adopted Sandage's (1972) modified
cosecant law, given by:

\small
\begin{eqnarray}
A_{\rm v}(b)_{\infty} & = & \nonumber \\
\left\{ \begin{array}{cl}
0.165 \times \frac{ \left( \tan 50^{\degr} - \tan b \right)}{\sin b} &
\mbox{if $10^{\degr} \leq |b| < 50^{\degr}$} \\
0 & \mbox {if $|b| \ge 50^{\degr}$}
\end{array}
\right.
\end{eqnarray}
\normalsize


where $A_{\rm v}(b)_{\infty}$ is the absorption, at infinity, that
goes into Eq. (4), after conversion to $E(B-V)(l,b)_{\infty}$ using
our adopted value of $R_{\rm v}$. Following Sandage, the above
expression is only applied for $10^{\degr} \leq |b| < 50^{\degr}$,
while it is assumed that $A_{\rm v}(b)_{\infty}=0$ at higher
latitudes. Sandage's law, based on an earlier model proposed by
McClure \& Crawford (1971), has been extensively criticized by de
Vaucoleurs \& Buta (1983). de Vaucoleurs \& Buta find, on the basis of
galaxy and galaxy clusters counts that a simple cosecant law (as
derived from a plane-parallel absorbing layer, Dufay 1954) applies all
the way to the poles, with a polar extinction of 0.2 magnitudes in
$A_{\rm B}$. de Vaucoleurs \& Buta also back their findings by
considering the surface brightness, the hydrogen-luminosity ratio, and
the integrated color indices of galaxies on the Second Reference
Catalogue of Bright Galaxies (de Vaucoleurs et al. 1976). de
Vaucoleurs \& Buta furthermore criticized Sandage's law on the basis
that it implies that the Sun is located in the common apex of two
opposite dust-free cones centered at the Galactic poles (see Fig.~1 on
de Vaucoleurs \& Buta). Sandage's law, on the other hand, agrees with
the very small color excesses found for stars and globular clusters at
high Galactic latitudes. In Fig.~1 we show a comparison between the de
Vaucoleurs \& Buta's and Sandage's law with the longitude-averaged run
of reddening from the BH maps. From this figure, it is clear that
Sandage's law is a better representation of the actual variation of
reddening as a function of Galactic latitude. The discrepancy between
Sandage's (zero-reddening) model and the BH maps for $|b| >
50^{\degr}$ is not relevant in this context, as most of the missing
points from the BH maps ocurr at smaller Galactic latitudes (except
for the region $b \le -60^{\degr}$ and $-130^{\degr} \le l \le
20^{\degr}$ where no data is available from the BH maps). Figure~1
also clearly shows that, at high Galactic latitudes, neither of the
two laws seem to be a good representation of the BH maps. Indeed,
Knude (1996) has recently found, from the color excesses derived from
Str\"omgren photometry for a complete sample of A3-G0 stars brighter
than $B=11.5$ and with $b>+70^{\degr}$, that the North Galactic Pole
exhibits a complex reddening distribution, with zonal variations (in
latitude and longitude) which can be represented more by what he calls
'a slightly frosted window'. This shows that {\it any} functional
representation for the total reddening, even at high Galactic
latitudes, has to be taken with caution. It is also interesting to
notice that Knude finds a very good agreement between the zero-point
on the BH maps (derived from colors of RR~Lyrae and individual stars
in globular clusters) and that derived by Knude's uvbyH$\beta$
photometry of individual stars towards {\it both} the south and north
galactic poles.

\begin{figure}
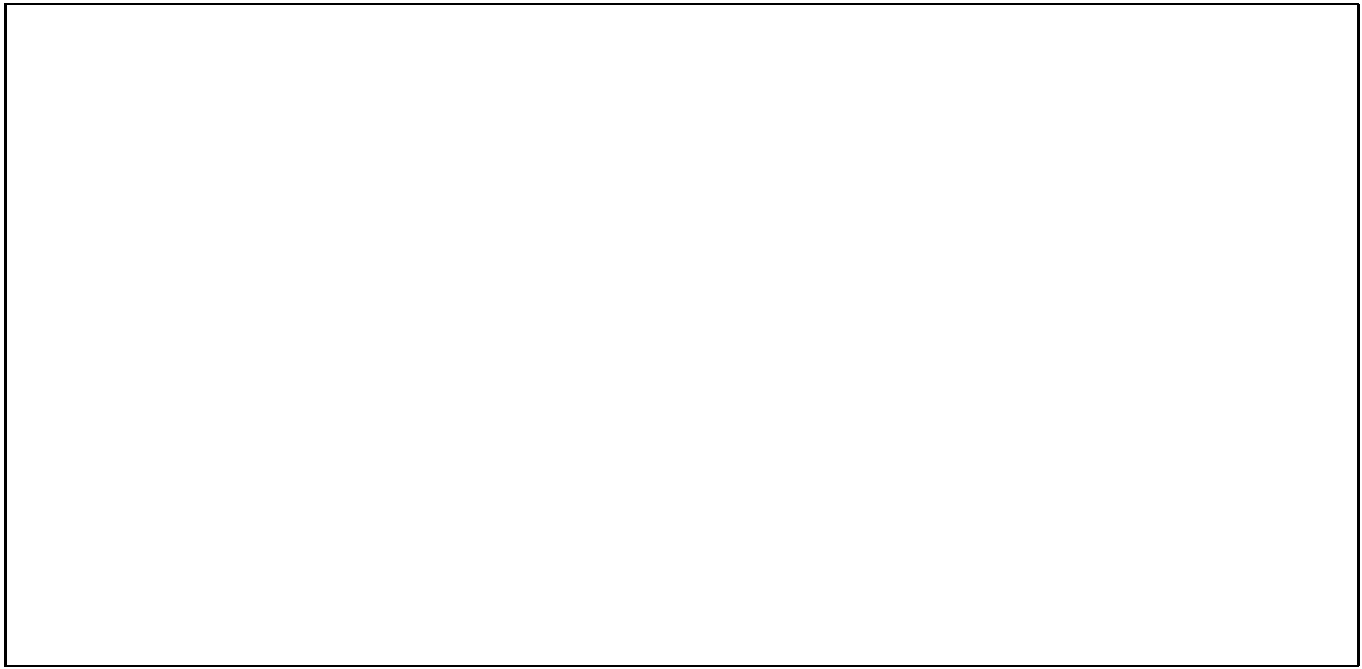

\picplace{8.8 cm}
\caption[]{Longitude-averaged reddening {\it vs.} Galactic latitude
from the Burstein \& Heiles maps (1982). Square symbols are for the
northern Galaxy ($b \ge +10^{\degr}$), circle symbols are for the
southern Galaxy($b \le -10^{\degr}$). Open symbols indicate mean
reddening with the corresponding error of the mean. Filled symbols
indicate the standard deviation. The solid line shows Sandage's (1972)
modified cosecant law, while the dashed lines indicate de Vaucoleurs
\& Buta's (1983) cosecant law, where we include the published range of
0.1 to 0.2 for $A_{\rm B}$ at the pole (Lu et al. 1992).}
\end{figure}

The adequacy of the interpolation scheme described above has been
demonstrated recently by M\'endez \& van Altena (1996), by performing
{\it simultaneous} fits to the magnitude and color counts for two
intermediate-latitude fields located at $(l,b) = (123^{\degr},
+22^{\degr})$ and $(287^{\degr}, +17^{\degr})$. The amount and
distribution of the reddening material is important for both the
starcounts (where reddening affects the total number of objects in a
given field-of-view) as well as the color-counts (where reddening
affects the color distribution of stars). The comparisons performed by
M\'endez \& van Altena indicate that, indeed, the exponential
approximation gives a very good fit to the star- and color-counts
towards these two fields having moderate extinction ($E({\rm
B-V})_{\infty} = 0.07 \mbox{ mag}$ and $0.05 \mbox{ mag}$ for the two
fields-of-view respectively). We should also point out that the use of
Eq. (4) for the case $Z_{\sun}=0$~pc is not new: it was proposed,
e.g., by van Herk (1965) in his study of the motions and space
distribution of RR~Lyrae stars, and it has been subsequently used in
different contexts where an accurate statistical determination of the
reddening is needed (e.g., Di~Benedetto \& Rabbia 1987, Fouqu\'e \&
Gieren 1996). Indeed, Di~Benedetto \& Rabbia (1987) have pointed out
that van Herk's model yields the same reddening estimates as those
derived from the method of color-excess described by Savage \& Mathis
(1979) when using a scale-height of 100~pc, close to our adopted value
(Sect. 2.2). Finally, J{\o}nch-S{\o}rensen (1994) has investigated the
reddening in six fields with $|b| \ge +14{\degr}$ up to distances of
10~kpc using uvbyH$\beta$ photometry. He finds quite a good agreement
in the run of reddening with distance compared to that predicted by a
simple exponential model similar to that of Eq.~(4) (see, e.g, his
Fig.~5), although he warns that simply adopting $E(B-V)(l,b)_{\infty}$
from the BH maps yields systematically lower reddening than observed,
except for the fields with the smaller reddening values. The largest
difference found among the six fields studied by J{\o}nch-S{\o}rensen
amounts to 0.15~mag in $E(B-V)$, with a mean difference for the six
regions of $0.081 \pm 0.054$~mag. This underestimation is partially a
volume-sampling effect, because of the coarse angular resolution of
the BH maps (derived from HI observations with a beam size of
$0.6{\degr}$) in comparison with the sizes of the fields studies by
J{\o}nch-S{\o}rensen (smaller than $0.06 \, \Box {\degr}$). Indeed, as
pointed out by Burstein \& Heiles (1982), their maps should be an
accurate representation of the reddening in the Galaxy in regions
where the variation of reddening with position is small over the
angular resolution of the maps. Evidently, for specific applications
(e.g, those involving small fields-of-view), the reddening should be
determined either from multi-color photometry or from the magnitude
and color counts, as done by M\'endez \& van Altena (1996). It is
interesting to notice that, in the study by M\'endez \& van Altena
(1996), involving two regions of about $0.5 \, \Box {\degr}$ each,
their derived reddening values are systematically {\it lower} than
those from the BH maps in an amount of 0.012~mag for their NGC~188
field, and 0.042~mag for their NGC~3680 field.

\subsection{Reddening at low Galactic latitudes}

For low-latitudes ($|b| < 10^{\degr}$), we have used published
reddening maps giving $E(B-V)$ {\it vs.} Heliocentric distance and
Galactic position. The main optical surveys devoted to the study of
the spatial distribution of interstellar reddening material in the
solar neighborhood have been those of FitzGerald (1968, 1974, 1987,
collectively called FG hereafter), FitzGerald \& Reed (1991), and
Neckel \& Klare (1980, NK hereafter), with a number of other more
restricted surveys (Deutschman et al. 1976, Forbes 1985, Pandey \&
Mahra 1987). In this study we have only included the reddening
determinations by FG and NK. FG's maps include a number of regions
with $|b| \approx 0^{\degr}$, while NK's maps are restricted to the
range $|b| < 7.6^{\degr}$, both studies cover the range $0^{\degr}
\leq l < 360^{\degr}$.

Since no computer-readable form existed for either FG's nor NK's maps,
we had to read-off their adopted mean reddening curves directly from
the figures in their publications (we should note that, given the
resolution of the maps presented by FG and NK, we estimate a
``reading'' uncertainty of about 0.03~mag in $E(B-V)$, while the
quoted errors for the reddening from the FG and NK maps are 0.015~mag
and 0.031 to 0.063~mag respectively). This involved the key-punching
of reddening maps for 75 lines-of-sight from FG's regions, and 316
lines-of-sight from NK's regions. Each line-of-sight consists itself
of ``node points'' at different Heliocentric distances. Therefore, our
low-latitude grid consists of 388 lines-of-sight (three lines-of-sight
were duplicated in both studies), with 1,263 independent node points
covering the whole Galactic plane with $|b| < 10^{\degr}$. The average
angular resolution in Galactic longitude is therefore around
$1^{\degr}$, but there are fluctuations in this number due to the
selective study of certain regions by both authors. Figure~2 shows the
distribution of node points with respect to the Sun, projected onto
the plane of the Galaxy (see also Fig. 4 in FitzGerald 1968, Fig. 9a
in NK, and Fig.~4 below).

\begin{figure}
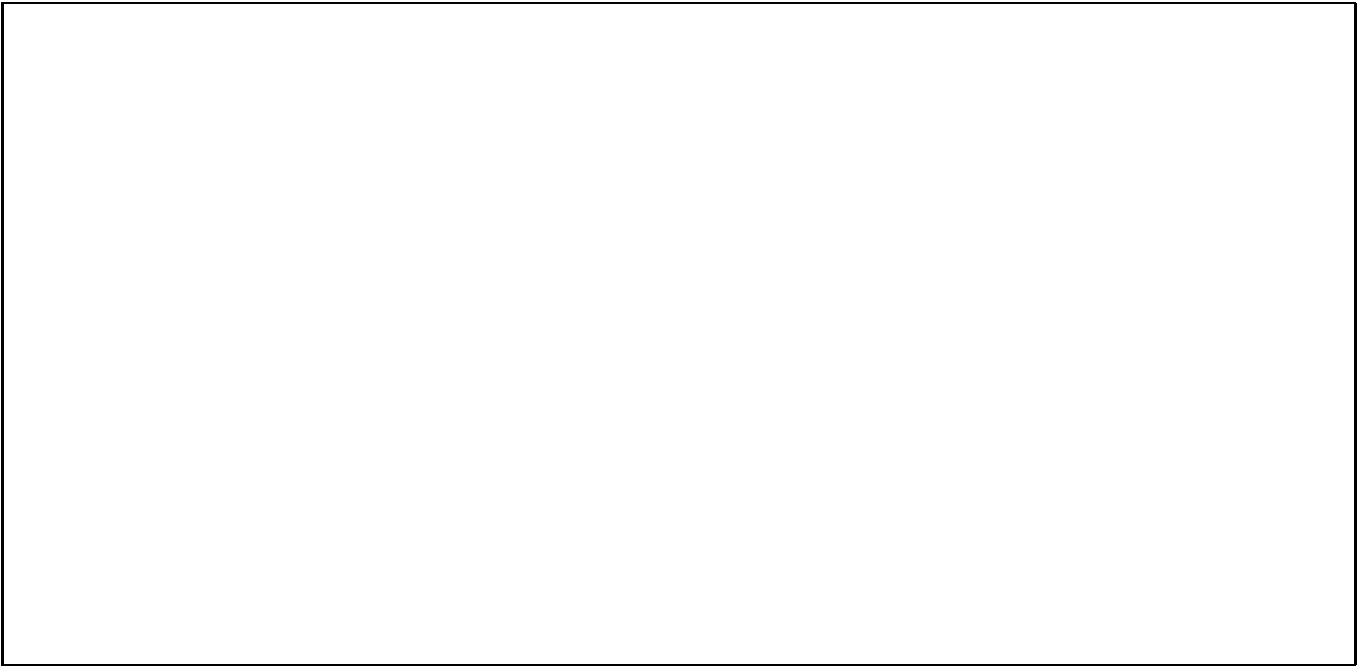

\picplace{8.8 cm}
\caption[]{Distribution of reddening node points on the plane of the
Galaxy with respect to the Sun. The V axis is positive in the
direction of Galactic rotation, while the U axis is positive toward
the Galactic center. The Sun is located at (0,0). The size of the
points is proportional to the reddening at that position, with the
smallest dots having $E(B-V)=0.03$, and the largest $E(B-V)=1.29
\mbox{ mag}$}
\end{figure}

In Fig.~3 we show a comparison between FG's and NK's maps for the
three directions in common between these two studies. NK's values of
$A_{\rm v}$ have been converted to $E(B-V)$ using his adopted value of
$R_{\rm v}= 3.1$. We can see that, at $l= 121^{\degr}$, there is a
good agreement between FG's and NK's values for $r< 1 \mbox{ kpc}$,
however there is an asymptotic difference in $E(B-V)$ of $0.4 \mbox{
mag}$ at $r \approx 3 \mbox{ kpc}$. The agreement at $l= 187^{\degr}$
is however very good at all distances ($r \leq 4 \mbox{
kpc}$). Finally, at $l= 245^{\degr}$, even though there is a good
agreement for $4< r \leq 6 \mbox{ kpc}$, FG's map shows a faster
increase at small distances ($r< 500 \mbox{ pc}$) and then it remains
constant up to $3.5 \mbox{ kpc}$, whereas NK's map indicates a
reddening approximately a factor of three bigger than FG's value.  For
these three directions we have chosen NK's values over those of
FG. The general agreement of these maps at small distances ($r< 1
\mbox{ kpc}$) is important, because most of the bright to intermediate
magnitude counts come from stars located at similar Heliocentric
distances. However, the noted discrepancies at large distances imply
that the adoption of these mean-extinction curves has to be taken with
caution, specially when trying to evaluate the reddening at large
Heliocentric distances.

\begin{figure}
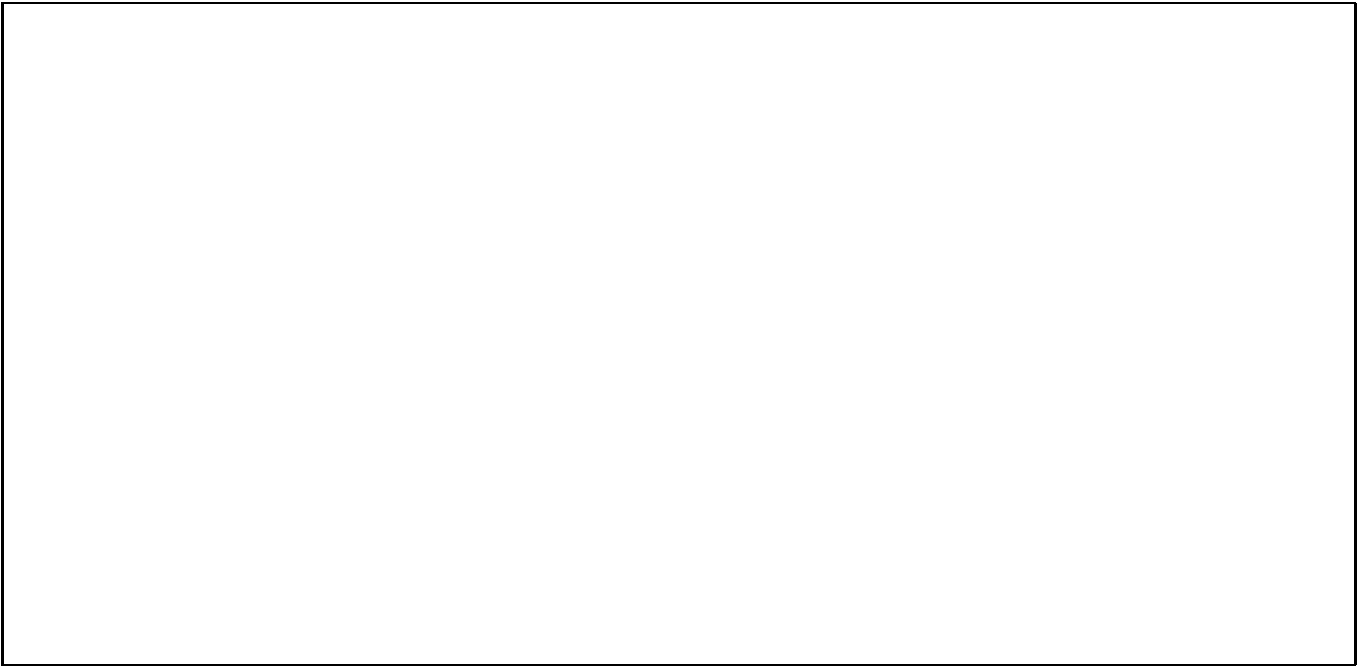

\picplace{8.8 cm}
\caption[]{Reddening values for three directions in common between
FG's and NK's reddening maps.}
\end{figure}

In the following section we describe the interpolation scheme used to
obtain reddening at positions other than those specified in FG's and
NK's reddening maps.

\subsection{Reddening interpolation at low Galactic latitudes}

\subsubsection{Generalities}

If the Galactic latitude of the point for which reddening is sought is
such that $|b|<10^{\degr}$, an interpolation scheme has been devised
that assumes that the distribution of reddening material presents both
a ``smooth'' component represented by an exponential distribution of
dust perpendicular to the Galactic plane (Sect. 2.2), and a ``clumpy''
component that represents the fragmentary nature of individual
molecular clouds along the line of sight, represented by the available
reddening maps. To accommodate for poorly sampled directions in the
plane, the BH reddening maps at $|b|= \pm 10^{\degr}$ are dynamically
incorporated in the interpolation. This also ensures a continuous
transition to higher latitudes.

If $(r,l,b)$ are the Heliocentric distance and the Galactic position
of the point for which reddening is sought (the ``target''), the
(combined) FG and NK reddening maps are searched for the two node
points in longitude that enclose the target longitude. Then, at those
two longitudes, the two latitudes that enclose the target latitude are
searched for. This search procedure may yield up to four node points
enclosing the target point. Then, at each enclosing node point, the
reddening at distance $r$ is interpolated linearly in distance from
the respective reddening maps. In this linear interpolation, if the
target distance is smaller than the first node point for a particular
reddening map, the value is interpolated assuming zero reddening at
zero distance and the reddening associated with the first distance
point in the reddening map. On the other hand, if the target distance
is greater than the last node point, the reddening is {\it
extrapolated}, as described in the following subsection.

In addition to the four nearby points, the reddening at Heliocentric
distance $r$, Galactic longitude $l$, and Galactic latitudes of $\pm
10^{\degr}$ is computed from the BH reddening maps. These reddening
values are used mainly for the intermediate region $5^{\degr} < |b| <
10^{\degr}$, where FG's and NK's reddening maps do not constrain the
reddening very well, and to provide a smooth transition to higher
Galactic latitudes.

\subsubsection{Reddening extrapolation}

We have attempted two different approaches to obtain the reddening at
distances beyond the last node point. First, we have tried assuming a
constant value of absorption per unit of distance (e.g., $1.5 \mbox{
mag/kpc}$ in $A_{\rm v}$), and using the last node point as a
zero-point, i.e.:

\begin{equation}
A_{\rm v}(r,l,b) = A_{\rm v}(r_l,l,b) + \delta A_{\rm v} \times \left( r - r_l \right)
\end{equation}

where $A_{\rm v}(r_l,l,b)$ is the reddening for the last distance node
located at $r_l$, and $\delta A_{\rm v}$ is the adopted absorption per
unit of distance. This extrapolation approach is similar to that
adopted in the extinction model by Arenou et al. (1992), we shall call
this Model-A. This model, which is quite common (see Sect 2.6.1), has
two main drawbacks. First, as we shall see below (Eq. 15), the
absorption in the solar vicinity has a Galactic longitude dependence
near the plane, and therefore it is not correct, in principle, to
assume a constant value for the differential reddening. However, since
this value is applied only after the last reddening point (typically
located at Heliocentric distances of 3~to 4~kpc (Fig.~2)), we can view
the term $\delta A_{\rm v}$ as a {\it residual reddening}, with a
minor effect on the bright starcounts discussed below, which are
dominated by stars located at distances of about 1~to 2~kpc (see Sect
3.3), whereas the term $A_{\rm v}(r_l,l,b)$ not only sets the
zero-point, but also determines the Galactic longitude dependence. The
second shortcoming of this approach is that, at large Heliocentric
distances and $b \neq 0^{\degr}$, this model assumes a constant value
of reddening for stars situated many (reddening) scale-heights from
the disk, leading therefore to an excess on the predicted
reddening. This problem has been alleviated in the model by Arenou et
al. (1992) by adopting zero differential reddening beyond a
Heliocentric distance such that the corresponding distance from the
Galactic plane is larger than the adopted scale-height of the
reddening material.

The second extrapolation scheme (Model-B) comes from insisting on the
functional form of the large-scale distribution of the reddening
material presented in Sect.~2.2, and from using, again, the last
reddening point, but this time as a scale factor (rather than as a
zero-point) for the extrapolation. If $E(r_l,l,b)$ is the last
reddening point, it is possible to show that the extrapolated
reddening to an Heliocentric distance $r$ (with $r> r_l)$ is given
by\footnote{Equation~(7) is only valid for $Z_{\sun} \ge 0$, but
similar expressions can be written for $Z_{\sun} < 0$.}:

\begin{eqnarray}
\frac {E(r,l,b)}{E(r_l,l,b)} = & \nonumber \\
\left\{ \begin{array}{ll}
\frac{1-{\rm exp }\left( -\frac{r\,   \sin b}{h_{\rm red}} \right)}
     {1-{\rm exp }\left( -\frac{r_l\, \sin b}{h_{\rm red}} \right)}
\; \mbox {if $b \ge 0^{\degr}$} \\

\frac {1- 2 \, {\rm exp }\left( \frac{Z_{\sun}}{h_{\rm red}} \right) +
{\rm exp }\left( \frac{2 \, Z_{\sun} + r \, \sin b}{h_{\rm red}}
\right) }
      {1- 2 \, {\rm exp }\left( \frac{Z_{\sun}}{h_{\rm red}} \right) +
{\rm exp }\left( \frac{2 \, Z_{\sun} + r_l \, \sin b}{h_{\rm red}}
\right) }
\; \mbox{if $b < 0^{\degr}$ and $r_l \, |\sin b| \ge Z_{\sun}$} \\

\frac{1-{\rm exp }\left( -\frac{r\,   \sin b}{h_{\rm red}} \right)}
     {1-{\rm exp }\left( -\frac{r_l\, \sin b}{h_{\rm red}} \right)}
\; \mbox {if $b < 0^{\degr}$ and $r \, |\sin b| \le Z_{\sun}$} \\

\frac {1- 2 \, {\rm exp }\left( \frac{Z_{\sun}}{h_{\rm red}} \right) +
{\rm exp }\left( \frac{2 \, Z_{\sun} + r \, \sin b}{h_{\rm red}}
\right) }
{1-{\rm exp }\left( -\frac{r_l\, \sin b}{h_{\rm red}} \right)}
\; \mbox {if $b < 0^{\degr}$, $r_l \, |\sin b| \le Z_{\sun}$, and 
$r \, |\sin b| \ge Z_{\sun}$} \\
\end{array}
\right.
\end{eqnarray}

We note that the above expression does not diverge for $b=0^{\degr}$,
in this case the extrapolated reddening is given by $E(r,l,b)=
E(r_l,l,b) \times r/r_l$. By comparing this simple expression with
Eq. (6), we see that a linear extrapolation is indeed a special case
of Eq. (7) for $b=0^{\degr}$ if we identify $\delta A_{\rm v}$ with
$A_{\rm v}(r_l,l,b)/r_l$. In Model-B, the Galactic longitude
dependency is, of course, built-into $E(r_l,l,b)$, and the decrease of
reddening beyond the scale-height of the obscuring material is fully
taken into account. The main problem with this model comes from the
fact that, since $E(r_l,l,b)$ acts as a scale-factor (rather than as a
zero-point as it was the case on Eq. (6)), the extrapolated reddening
is extremely sensitive to uncertainties in the last reddening point
which, being located at large Heliocentric distances, is subject to a
greater uncertainty than the node points located at smaller
Heliocentric distances, a point clearly demonstrated by Fig.~3 (see
also discussion at end of Sect.~2.4). In Sects. 2.6.1 and 2.6.2 we
fully compare the results of adopting either model, and its impact on
the predicted reddening.

\subsubsection{Weighting scheme \& interpolation}

Once the reddening of the four adjacent node points has been found,
the reddening at the target is found through a {\it weighted} mean of
the reddening from those four node points, plus the scaled reddening
values from the BH maps. If $w_i$ (i=1...4) are the assigned weights
to the four enclosing reddening points with reddening $E_i$, and
$w_{\rm p}$, and $w_{\rm n}$ are the weights assigned to the BH
reddening values at $b=+10^{\degr}$ and $b=-10^{\degr}$ respectively,
then the interpolated reddening is computed via:

\begin{equation}
E(r,l,b) = \frac 
{ \sum_{i \leq 4} w_i \times E_i + w_{\rm p} \times E_{\rm p} + w_{\rm n} \times E_{\rm n}}
{\sum_{i \leq 4} w_i + w_{\rm p} + w_{\rm n}}
\end{equation}

where $E_{\rm p}$ and $E_{\rm n}$ are the values derived from the BH
reddening maps (Eq. (4)), at $b=+10^{\degr}$ and $b=-10^{\degr}$
respectively.

We have incorporated the clumpy nature of the absorbing material, as
well as its large-scale properties, through a proper selection of the
weights specified in Eq.~(8). The weights $w_i(r,l,b)$ are given by:

\begin{eqnarray}
w_i(r,l,b) & = & \tilde w_i(r,l,b) \times f(\delta S_i) \\
\delta S_i & = & \sqrt{2 \, \left( 1 - \cos \theta_i \right)}
\times r \\
\cos \theta_i & = & \cos b \, \cos b_i \, \cos l \, \cos l_i + \nonumber \\
& & \cos b \, \cos b_i \, \sin l \, \sin l_i + \sin b \, \sin b_i
\end{eqnarray}

where $(l_i,b_i)$ are the Galactic longitude and latitude of the i-th
node point, $\tilde w_i(r,l,b)$ are the weights that describe the
large-scale variation of the reddening material, $f(\delta S_i)$ is a
function that considers the patchy distribution of reddening material,
and $\delta S_i$ and $\theta_i$ are the linear and angular distances
between the corresponding node points and the target respectively. The
weights $\tilde w_i(r,l,b)$ have two components, a latitude component,
$w^{\rm b}_i(r,b)$, and a longitude component, $w^{\rm l}_i(r,l)$,
both of which are described below. The large-scale component of the
weights is thus given by $\tilde w_i(r,l,b) = w^{\rm b}_i(r,b) \times
w^{\rm l}_i(r,l)$.

The weights $w^{\rm b}_i(r,b)$ are computed by taking the ratio of the
expected value for the reddening at position $(r,l,b)$, and the
expected value at the respective node position. Effectively this ratio
scales the (observed) reddening value at the node point to that at the
target location. This is achieved by using our assumed density law for
the dust particles (Eq. (2)) integrated over the appropriate
line-of-sight (Eq. (1)). The weights $w^{\rm b}_i(r,b)$ are thus given
by:

\begin{equation}
w^{\rm b}_i(r,b) = \frac{\Omega(r,b)}{\Omega(r,b_i)}
\end{equation}

where the function $\Omega(r,b)$ is computed as\footnote{Equation (13)
is valid only for $Z_{\sun} \ge 0$, but similar expressions can be
written for $Z_{\sun} < 0$.}:

\begin{eqnarray}
\Omega(r,b) = & \nonumber \\
\left\{ \begin{array}{ll}
\frac{1-{\rm exp }\left( -\frac{r\, \sin b}{h_{\rm red}} \right)} {\sin b} 
\; \mbox {if $b \ge 0^{\degr}$} \\
\frac{1-{\rm exp }\left( {-\frac{r\, \sin b}{h_{\rm red}}} \right)}
{\sin b} \; \mbox{if $b < 0^{\degr}$ and $r \, |\sin b| \le Z_{\sun}$} \\
\frac {1- 2 \, {\rm exp }\left( \frac{Z_{\sun}}{h_{\rm red}} \right) +
{\rm exp }\left( \frac{2 \, Z_{\sun} + r \, \sin b}{h_{\rm red}} \right) }
{\sin b} \;
\mbox{if $b < 0^{\degr}$ and $r \, |\sin b| > Z_{\sun}$}
\end{array}
\right.
\end{eqnarray}

For example, if $Z_{\sun}= 0$, the weights $w^{\rm b}_i(r,b)$ (for
{\it any} value of $r$, $b$, and $b_i$) take the very simple
expression:

\begin{equation}
w^{\rm b}_i(r,b) = \frac
{ \left( 1 - {\rm exp }\left( -\frac{r\, |\sin b|}{h_{\rm red}} \right) \right) }
{ \left( 1-{\rm exp }\left( -\frac{r\,|\sin b_i|}{h_{\rm red}} \right) \right) } \times
\frac{|\sin b_i|}{|\sin b|}
\end{equation}

Note also that, for $b=0^{\degr}$, $\Omega(r,b)$ does not diverge, but
it approaches the ratio $r/h_{\rm red}$ which corresponds to a linear
increase of reddening with distance (see Eq. (6), and comments
following Eq. (7)). Indeed, this is the only possible asymptotic
solution for an exponential layer, as $b$ goes to zero.

In addition to the Galactic-latitude (or height above the plane)
dependence of the weights, we have implemented also a longitude
dependence. From a study of classical Cepheids, Fernie (1968) found
that the reddening exhibits a cyclic (sinusoidal) variation with
Galactic longitude at $b=0^{\degr}$, while Pandey \& Mahra (1987)
found a similar behavior for $|b|<10^{\degr}$ from a sample of 462
open clusters with known color excesses and (photometric)
distances. For the cyclic variation represented by $\delta A_{\rm
v}(l) = \delta A_{\rm v}^{\rm o} + \delta A_{\rm v}^{\rm l} \, \sin
(l+l_{\rm o})$, Fernie (1968) found $\delta A_{\rm v}^{\rm o}= 0.90
\pm 0.04 \mbox{ mag/kpc}$, $\delta A_{\rm v}^{\rm l} = 0.28 \pm 0.06
\mbox{ mag/kpc}$, and $l_{\rm o}= 41^{\degr} \pm 13^{\degr}$ from
Cepheids located at Heliocentric distances closer than 6 kpc. Pandey
\& Mahra (1987) found a maximum absorption at $l \approx 50^{\degr}$,
and a minimum absorption at $l \approx 230^{\degr}$, for $r< 2 \mbox{
kpc}$ (where their sample of open clusters is not extremely biased),
in accordance with Fernie's value for $l_{\rm o}$. Lyng{\aa} (1979,
1982) also found a maximum average interstellar absorption toward $l
\approx 50^{\degr}$, based on a study of B stars, classical Cepheids,
and open clusters.

Based on Fernie's results, our weights in Galactic longitude,
$w^{\rm l}_i(r,l)$, have been defined as:

\begin{equation}
w^{\rm l}_i(r,l) = \left\{ \begin{array}{cl}
\frac
{\delta A_{\rm v}^{\rm o} + \delta A_{\rm v}^{\rm l} \, \sin (l+l_{\rm o})}
{\delta A_{\rm v}^{\rm o} + \delta A_{\rm v}^{\rm l} \, \sin (l_i+l_{\rm o})}
& \mbox{if $r \leq 6$ kpc} \\
1 & \mbox {otherwise}
\end{array}
\right.
\end{equation}

Since Fernie's relationship seems to pertain only to the local
distribution of reddening material, the weights given in Eq.~(15) are
only applied for distances smaller than 6 kpc.

As for the function $f(\delta S)$ in Eq. (9), we have tried different
functional forms, and we have chosen the one that minimizes the
reddening residuals of the observed {\it vs.} predicted node point in
a self-consistent way, as described in Sect. 2.6. The purpose of
introducing this function has been to take into account the discrete
nature of the individual clouds and the patchy distribution of the
interstellar dust by weighting the node points according to its
proximity to the target location.

We should note that, since the BH reddening maps are evaluated at $(l,
\pm 10^{\degr}$), then Eq.~(11) becomes $\cos \theta_{\rm pn} = \cos b
\, \cos 10^{\degr} \pm \sin b \, \sin 10^{\degr}$, where the + sign
applies to the node point with reddening $E_{\rm p}$ (with weight
$w_{\rm p}$), while the - sign applies to the node point with
reddening $E_{\rm n}$ (with weight $w_{\rm n}$) from the BH
maps. Also, for the BH reddening values we do not apply any Galactic
longitude variation component to the weights (which are applicable
only to $b \sim 0^{\degr}$), but only the latitude dependence
(Eq. (12)).

Overshoot and undershoot in latitude in those cases where the target
latitude is bigger or smaller respectively than the biggest and
smallest latitude on the reddening arrays is taken care of by
considering only the closest latitude point with the proper
weight. Therefore the interpolation is set up to work with a maximum
of six points (four node points from the low-latitude reddening maps,
and two node points from the BH maps) to a minimum of four points
(two node points from the low-latitude reddening maps, and two node
points from the BH maps). Note that in cases of over/undershoot we
extrapolate in latitude using our exponential reddening model
described above. Since reddening data is available in the whole range
$0^{\degr} \leq l < 360^{\degr}$, there is never over/undershoot in
longitude.

It should be noticed that, with the weighting scheme described above,
the role of the BH maps, even at low Galactic latitudes, is quite
important. For example, for a target point at $(r,l,b)=(1 \, {\rm
kpc},90^{\degr},+1^{\degr})$ with a node point at
$(l,b)=(90^{\degr},0^{\degr})$, the weights applied to the node point
and the values from the BH maps at $\pm 10^{\degr}$ are 0.91, 1.84,
and 1.37 correspondingly, where we have adopted for $f(\delta S)$ an
inverse-linear relationship (see Sect.~2.6), $Z_{\sun}=+7$~pc, and
$h_{\rm red}= 110$~pc. In cases where the target location is enclosed
by four node points from the FG or NK maps, one might not use the BH
maps, and could perform an interpolation using only those node
points. However, due to the relatively poor angular sampling of the
current maps (see Sect.~2.4 and Fig.~2) it seldom occurs that a target
is surrounded by four node points, and therefore this is not a problem
for the current implementation of the model, but it could become
important as the density of node points increases through the
incorporation of more reddening maps in our database.

A simpler approach, but similar to that of Eq. (8), has been used
recently by Layden et al. (1996) in the context of statistical
parallaxes for RR~Lyrae stars. Indeed, this is yet another possible
use of our starcounts \& kinematic model, which we intend to pursue in
the future.

\subsection{Tuning the low Galactic latitude interpolation}

The following procedure has been adopted to estimate the expected
uncertainty in the interpolated reddening values: the reddening has
been computed at each node point for a given longitude (a ``longitude
slice'') from the two adjacent node points in longitude, and the
root-mean-square (rms) residual between the predicted value and the
observed one has been obtained for each of those longitude slices. Our
reddening maps have 276 independent longitude slices. Each longitude
slice may contain more than one latitude node (although it has,
typically, only one latitude point at $b \sim 0^{\degr}$) and, at each
latitude, there may be node points at more than one Heliocentric
distance. This scheme is similar to the way in which reddening is
computed for a line-of-sight integration in our starcounts model
(Sect. 3.2), and therefore it provides a realistic description of the
accuracy of our reddening model.

Using the above procedure we can examine the residuals between our
model predictions and the observed values at each node point in the
reddening maps in a self-consistent manner. This allows us to
``tune-up'' the adopted parameters, as well as the exact distance
weighting scheme in the reddening model (the term $f(\delta S)$ in
Eq. (9)), so as to minimize the scatter between the predicted and
observed values. In the following two subsections we deal, separately,
with the Models-A and~B described in subsection~2.5.2 (Eqs.~(6)
and~(7)).

\subsubsection{Tuning Model-A}

Figure 4 shows the residuals (in the sense $E(B-V)_{\rm model} -
E(B-V)_{\rm node}$) as a function of Galactic latitude, longitude, and
Heliocentric distance for Model-A. Figures 4a-b provide and indication
of the angular sampling with the reddening maps implemented in the
current version of our reddening model, it also shows that we do not
have any obvious systematic trend in the predicted residuals as a
function of either Galactic latitude or longitude. Figure 4c shows, on
the other hand, an upward trend in the residuals {\it vs.} distance,
particularly for $r > 2 \mbox{ kpc}$. This is caused by a value of
$\delta A_{\rm v}$ beyond the last node point that is too large. The
value of $\delta A_{\rm v}$ used in Fig. 4c was $1.5 \mbox{
mag/kpc}$. Table 1 indicates the dispersion in the reddening model as
a function of $\delta A_{\rm v}$. The best value for $\delta A_{\rm
v}$, obtained from a minimization of the residuals as a function of
this parameter, gives $\delta A_{\rm v} = 0.5 \mbox{ mag/kpc}$, which
for a value of $R_{\rm v} = 3.2$ corresponds to only $0.16 \mbox{
mag/kpc}$ of reddening beyond $2 \mbox{ kpc}$. This value is
considerably {\it smaller} than previous determinations of $\delta
A_{\rm v}$, which yielded values closer to $1.5 \mbox{ mag/kpc}$. For
example, Van Herk (1965) adopted a value of 1.9~mag/kpc (for $|b|>
40^{\degr}$), and Di Benedetto \& Rabbia (1987) a value of
1.4~mag/kpc. On the other hand, Lyng{\aa} (1982)
adopted a value of $0.75 \mbox{ mag/kpc}$ from an statistical study of
a sample of open clusters. More recently, Blackwell et al. (1990) have
discussed the evidence for even smaller values of $\delta A_{\rm v}$
from the {\it local} (i.e., $r< 300$~pc) distribution of reddening
material around the Sun, adopting in the end an absorption of
0.6~mag/kpc, quite close to our derived value. We should note,
however, that those determinations were concerned with {\it local}
reddening, while our value is determined mainly from the more distant
node points (further than 2 kpc from the Sun). Figure~4d shows the
residuals {\it vs.} distance for $\delta A_{\rm v} = 0.5 \mbox{
mag/kpc}$ (Figs. 4a-b were produced with this value of $\delta A_{\rm
v}$ as well). The residuals {\it vs.} distance vary slowly with
$\delta A_{\rm v}$ and, as a result, our minimum is rather
shallow. This is because a large fraction of the residual is produced
by the nearby points that are not sensitive to the actual value of
$\delta A_{\rm v}$. However, the effect of different values of $\delta
A_{\rm v}$ at large Heliocentric distances is obvious in Figs. 4c-d.

\begin{figure}
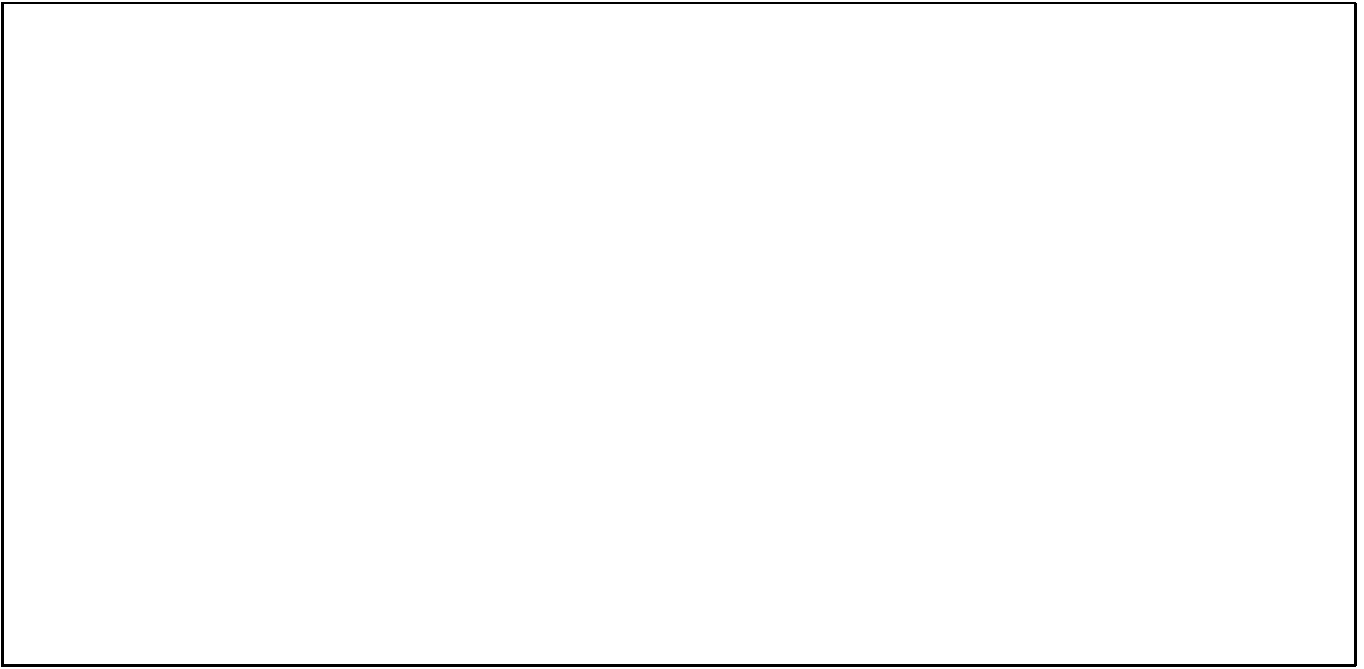

\picplace{8.8 cm}
\caption[]{Reddening residuals (in the sense $E(B-V)_{\rm model} -
E(B-V)_{\rm node}$) {\it vs.} Galactic position for Model-A. Panels
(a) and (b) indicate the reddening residuals {\it vs.} Galactic
longitude and latitude respectively. Panels (c) and (d) show the
reddening residuals {\it vs.} Heliocentric distance. In panels (a),
(b) and (d) we have used $\delta A_{\rm v} = 0.5 \mbox{ mag/kpc}$,
while in panel (c) we have used $\delta A_{\rm v} = 1.5 \mbox{
mag/kpc}$ (see text for explanation).}
\end{figure}

Table 1 also indicates the effect of changing $R_{\rm v}$ for the case
when $\delta A_{\rm v}$ is fixed to 0.5~mag/kpc. The effect on the
derived dispersion is minimal, even when changing $R_{\rm v}$ by as
much as 13\%. In terms of residuals, there is very little difference
when changing $R_{\rm v}$ for $r <$~3 -- 4~kpc. The trend we
find is that a larger $R_{\rm v}$ predicts somewhat smaller model
reddenings (i.e., a more negative residual) for $r > 4 \mbox{ kpc}$,
thus allowing for a larger value of $\delta A_{\rm v}$. This is
expected, since the quantity of interest is actually the differential
reddening, which is the ratio between $\delta A_{\rm v}$ and $R_{\rm
v}$. The lack of sensitivity of the reddening model values to the
exact value of $R_{\rm v}$ is important in view of the rather large
sensitivity of our model starcounts to this parameter, as shown in
Sect. 3.2. Figures 5a-b show the residuals {\it vs.} Heliocentric
distance for these two extreme values of $R_{\rm v}$, it is evident
that the differences in the distribution of residuals is minimal.

We have also varied the values of $h_{\rm red}$ and $Z_{\sun}$ within
considerable limits. Since these parameters enter only in the
``smooth'' component of the weights, as described in Sect. 2.5.3, they
have a rather small impact on the computed residuals. Table~1 lists
our results for some extreme cases of these parameters, along with the
computed dispersion. We find no clear trend in the reddening residuals
{\it vs.} Galactic position or Heliocentric distance for different
values of $Z_{\sun}$. For $h_{\rm red}= 50 \mbox{ pc}$ there seems to
be a small downward trend in the range $1 < r < 2 \mbox{ kpc}$ and
then a flattening at larger distances, while for $h_{\rm red}= 200
\mbox{ pc}$ there seems to be a small upward trend for $r > 3 \mbox{
kpc}$, although these features are not very pronounced (see
Figs. 5c-d). As for the dependence on $Z_{\sun}$ (Figs. 5e-f), we find
very little dependence on this parameter, with an indication of a
slight upward trend for $r > 4 \mbox{ kpc}$ when $Z_{\sun}= 40 \mbox{
pc}$. As a consequence, we can not constrain these parameters from
these data alone. It is not surprising that this test is not
sensitive, in particular, to $h_{\rm red}$ because our node points are
mostly located at $b \sim 0^{\degr}$, and therefore the dependence on
height above/below the Galactic plane is greatly reduced. Therefore,
in what follows, we have adopted the values of $h_{\rm red} = 110
\mbox{ pc}$ and $Z_{\sun} = +7 \mbox{ pc}$, as indicated in Sect. 2.2,
and in M\'endez \& van Altena (1996) respectively.

\begin{figure}
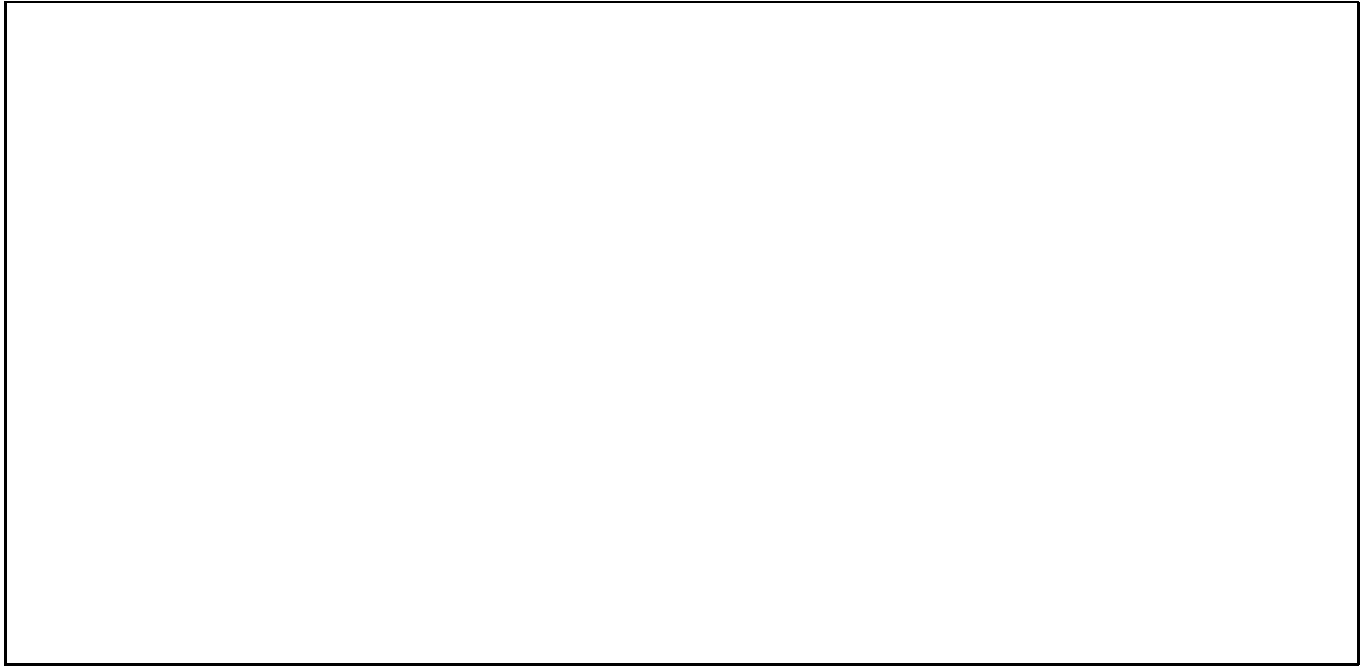

\picplace{8.8 cm}
\caption[]{Reddening residuals {\it vs.} Heliocentric distance. Panels
(a) and (b) show the reddening residuals for $R_{\rm v}=2.8$ and
$R_{\rm v}=3.6$ respectively. Panels (c) and (d) show the reddening
residuals for $h_{\rm red}=50 \mbox{ pc}$ and $h_{\rm red}=200 \mbox{
pc}$ respectively. Finally, panels (e) and (f) show the reddening
residuals for $Z_{\sun}=0 \mbox{ pc}$ and $Z_{\sun}= 40 \mbox{ pc}$
respectively (see Table 1).}
\end{figure}

As for the distance-weighting function $f(\delta S)$ in Eq. (9), we
have tried different functional forms, and have chosen the one that
minimizes the overall scatter. Table~1 indicates the mean residual
values for different functional forms, adopting $\delta A_{\rm
v}=0.5$~mag/kpc. We have tried inverse-linear, inverse-square,
inverse-cubic, and exponentially-decaying weights (see Fig. 6). In all
cases (except the exponentially-decaying weight) we have used a
softening parameter of 1 pc, the exponential-decay was given a
scale-length of 1 pc as well. The best result is obtained with the
inverse-linear weight (indeed, Models~A-1 through~A-8 in Table~1 were
performed precisely with this weighting scheme, with the best
inverse-linear solution being Model~A-2). The inverse-square,
inverse-cubic, and exponentially-decaying weights exhibit an
intrinsically larger scatter as a function of distance, and they also
exhibit an upward trend in the residuals for distances larger than 4
kpc. Since for these models we have already adopted a small value of
$\delta A_{\rm v}$ it seems unlikely that this upward trend is the
result of an inadequate value for this parameter, but rather it shows
the inadequacy of these weights. We have also tried no-distance
weights at all (i.e., $f(\delta S)=1$). In this case (see Table 1,
Model~A-12), the residuals are larger than for the inverse-distance
weight and there is a small {\it downward} trend in the residuals {\it
vs.}  distance. By increasing the value of $\delta A_{\rm v}$, as
indicated in Table 1 (Models~A-13 \&~A-14), the residuals do decrease
slightly, but are still larger than for an inverse-linear weight. For
values of $\delta A_{\rm v} > 1.0 \mbox{ mag/kpc}$ the residuals
wiggle, going downwards in the range $2 \mbox{ kpc} < r < 3 \mbox{
kpc}$ and then upwards for $r > 3 \mbox{ kpc}$. Also, as it can be
seen from Table 1, the mean value of the residual is not zero,
indicating a systematic offset between the model predictions and the
observed values. This offset is not present when using {\it any}
distance-weighting.

\begin{figure}
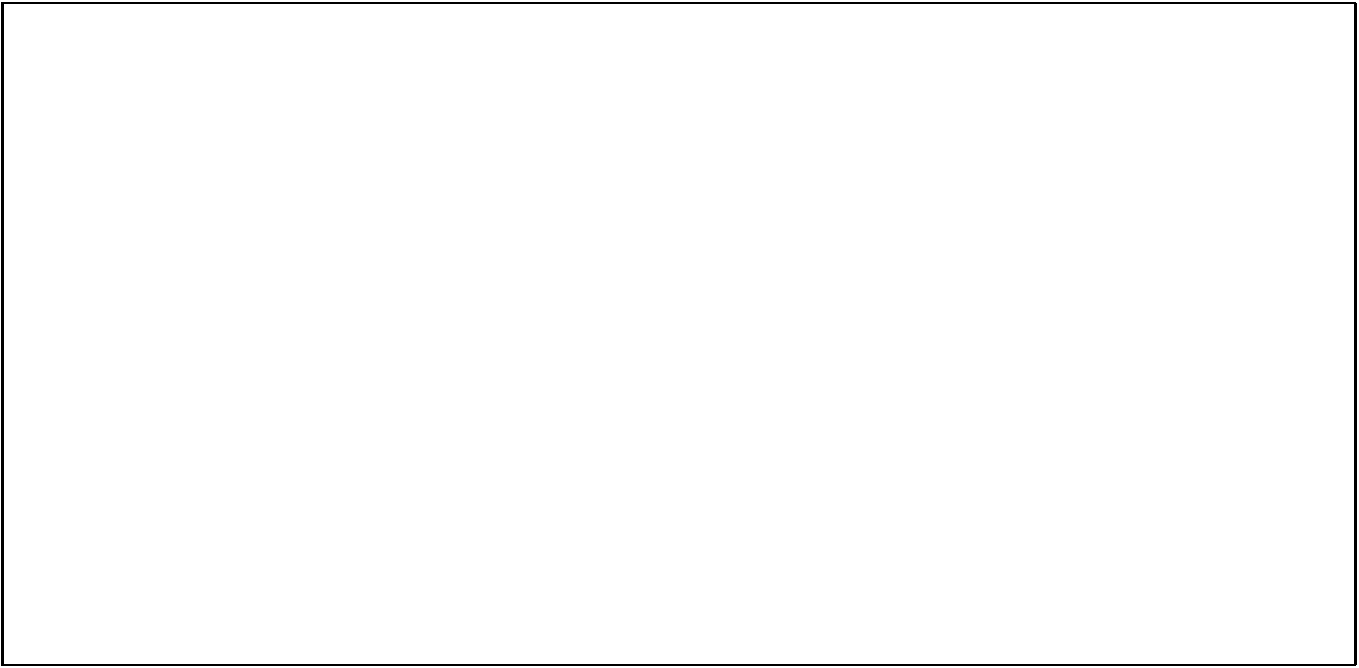

\picplace{8.8 cm}
\caption[]{Reddening residuals {\it vs.} Heliocentric distance. Panels
(a) and (b) show the reddening residuals for an inverse-square and an
inverse-cubic weight respectively. Panel (c) shows the reddening
residuals for an exponentially-decaying weight. Panels (d), (e) and
(f) show the reddening residuals for no distance weight with $\delta
A_{\rm v} =0.5$, $1.1$, and $1.5 \mbox{ mag/kpc}$ respectively (see
text for details).}
\end{figure}

%
%

\begin{table}
\caption[]{Mean reddening residual \& dispersion for Model-A.}
\begin{flushleft}
\begin{tabular}{cccc}
\hline\noalign{\smallskip}
Model & $<Res>$ & $\sigma_{Res}$ & Comments \\
\noalign{\smallskip}
\hline\noalign{\smallskip}
A-1 & -0.03  & 0.25 & $\delta A_{\rm v} = 1.5  \mbox{ mag/kpc}$ \\
A-2 & -0.07  & 0.23 & $\delta A_{\rm v} = 0.5  \mbox{ mag/kpc}$ \\
A-3 & -0.07  & 0.23 & $R_{\rm v} = 2.8 \mbox{ mag}$ \\
A-4 & -0.07  & 0.23 & $R_{\rm v} = 3.6 \mbox{ mag}$ \\
A-5 & -0.08  & 0.24 & $h_{\rm red} = 50 \mbox{ pc}$ \\
A-6 & -0.06  & 0.23 & $h_{\rm red} = 200 \mbox{ pc}$ \\
A-7 & -0.07  & 0.23 & $Z_{\sun} = 0  \mbox{ pc}$ \\
A-8 & -0.06  & 0.23 & $Z_{\sun} = 40 \mbox{ pc}$ \\
A-9 & -0.02  & 0.24 & inverse-square weight \\
A-10 & +0.01  & 0.25 & inverse-cubic weight \\
A-11 & +0.03  & 0.28 & exponentially-decaying weight \\
A-12 & -0.15 & 0.27 & no weight, $\delta A_{\rm v} = 0.5  \mbox{ mag/kpc}$ \\
A-13 & -0.14 & 0.23 & no weight, $\delta A_{\rm v} = 1.1  \mbox{ mag/kpc}$ \\
A-14 & -0.13 & 0.23 & no weight, $\delta A_{\rm v} = 1.5  \mbox{ mag/kpc}$ \\
\noalign{\smallskip}
\hline
\end{tabular}
\end{flushleft}
\end{table}

From the results presented in the preceding paragraph we conclude that
the best weighting scheme is that of an inverse-linear weight. It is
interesting to notice that there might be a physical reason for this:
In an extensive study of the Rosette and Maddalena-Thaddeus Molecular
Clouds (the latter discovered by Maddalena \& Thaddeus 1985), Williams
(1995, see also Williams et al. 1995) has shown that the
(line-of-sight) {\it column density} of $^{13}$CO in individual clumps
can be fit by a power law of the type $N_{\rm{13_{CO}}} \sim 1/(1 +
r/0.7 \mbox{ pc})$, close to what one might expect for the projection
of an isothermal clump with $\rho (r) \sim 1/r^2$, where $\rho (r)$ is
the volume density in the clump. Since our reddening values are
directly related to the column density of absorbing material, the fact
that our best weighting scheme corresponds to that found by Williams
for isolated clouds indicates that the overall reddening is, {\it on
the mean}, caused by clouds with the same density profiles as those
studied by Williams, with a typical core-radius (our ``softening
parameter'') close to 1 pc. We should also point out that Williams'
results hold for pressure-bound, gravitationally-bound, and
star-forming clumps, with minor differences in the slopes of the
projected column-density for these different clouds.

The histogram distribution of the (slice averaged) residuals is shown
in Fig. 7. The mean of the rms value for all longitude slices turns
out to be $0.23 \mbox{ mag}$ in $E(B-V)$ (Table 1, Model~A-2), with
few outliers, as evidenced by the ratio of the mean of the rms values
to the mean of the absolute dispersions ($0.20 \mbox{ mag}$). The
ratio of these two values should approach asymptotically to 1.25
(Meyer 1975) for a Gaussian function, as it is indeed the case for our
model. This rms value has to be regarded as a measure of the intrinsic
clumpiness of the reddening in the reddening maps rather than as an
estimation of the correctness of the weighting or interpolation scheme
itself. We would expect, e.g., that this rms value will decrease as
the node points become denser through the incorporation of more
extensive reddening surveys. Our reddening model seems to have a
substantially larger scatter than that of the extinction model by
Arenou et al. (1992), who report a standard deviation of 0.14 in
$E(B-V)$ by comparing predicted to observed colors for stars
observed with the Hipparcos star mappers. Our comparisons to
starcounts at low-Galactic latitudes seem to indicate, however, that
our model gives a better match than the Arenou et al. model to the
observed starcounts (see Sect. 3.2).

\begin{figure}
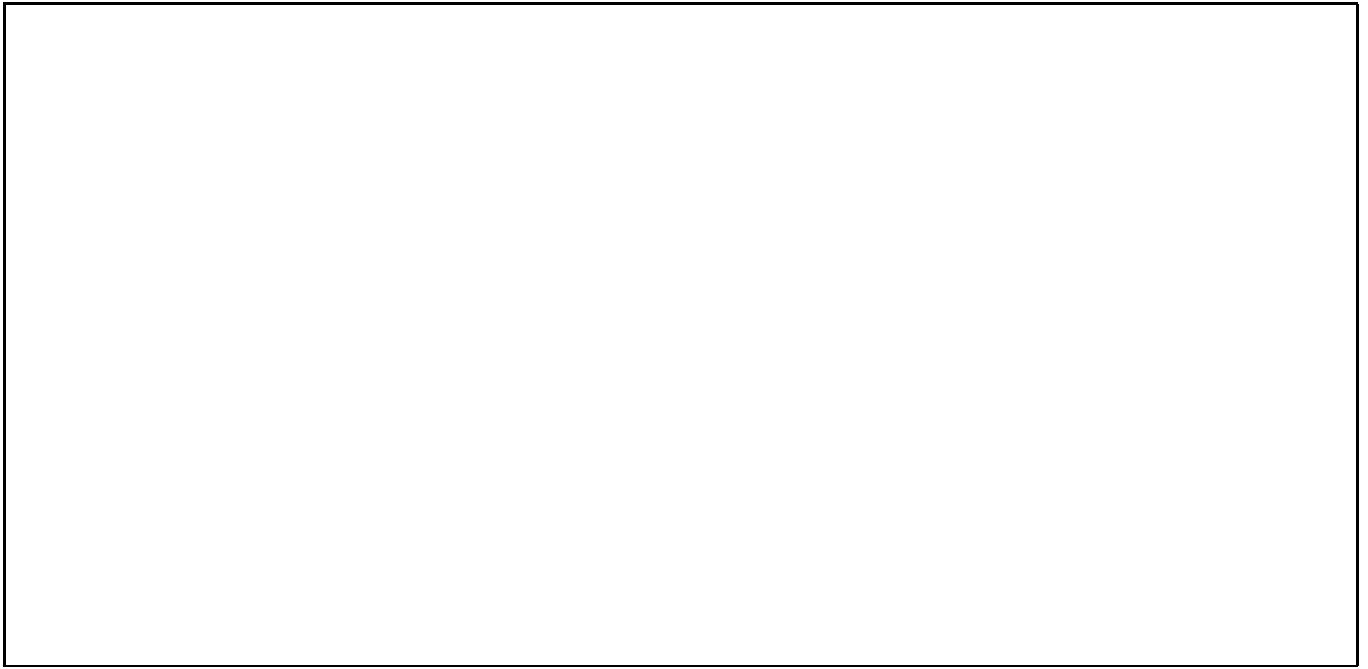

\picplace{8.8 cm}
\caption[]{Histogram of reddening residuals. The estimated rms value
for the dispersion is $0.23 \mbox{ mag}$ in $E(B-V)$.}
\end{figure}

\subsubsection{Tuning Model-B}

We follow the same analysis as for Model-A, noticing that in Model-B
we have one less degree of freedom, since $\delta A_{\rm v}$ in
Eq.~(6) is now determined by the last reddening point (see discussion
on Sect.~2.5.2).

In Table~2 we indicate the mean residual difference and the residual
dispersion between the model and the predictions. We find the same
trends as for Model-A, indicating that our tuning is nearly
independent of the more distant (and, therefore, uncertain) node
points (with the possible exception of $\delta A_{\rm v}$ which is
determined mainly by the more distant node points, see the previous
section). In particular, the dependency on the distance-weighting
function is much stronger in this case, with a clear indication that
the best weighting is that of an inverse-linear relation (compare
model B-1 to models B-8, B-9, B-10 \& B-11). The overall dispersion of
our best Model-B (B-1, with 0.25~mag) is only slightly larger than
that for our best Model-A (A-2 with 0.23~mag).

In Fig.~8 we show the residuals as a function of Galactic position
and Heliocentric distance for model B-1. No curvature on the residuals
as a function of distance is seen (compare with Fig.~~4c), which
shows the relative advantage of using the more self-consistent
extrapolation to larger distances in Model-B rather than the adoption
of a fixed reddening slope beyond the last node point, as done in
Model-A. The histogram of residual values is similar to that of
Model-A, and it is not shown here.

\begin{table}
\caption[]{Mean reddening residual \& dispersion for Model-B.}
\begin{flushleft}
\begin{tabular}{cccc}
\hline\noalign{\smallskip}
Model & $<Res>$ & $\sigma_{Res}$ & Comments \\
\noalign{\smallskip}
\hline\noalign{\smallskip}
B-1 &  -0.05  & 0.25 & standard \\
B-2 &  -0.05  & 0.25 & $R_{\rm v} = 2.8 \mbox{ mag}$ \\
B-3 &  -0.05  & 0.25 & $R_{\rm v} = 3.6 \mbox{ mag}$ \\
B-4 &  -0.08  & 0.24 & $h_{\rm red} = 50 \mbox{ pc}$ \\
B-5 &  -0.04  & 0.26 & $h_{\rm red} = 200 \mbox{ pc}$ \\
B-6 &  -0.06  & 0.25 & $Z_{\sun} = 0  \mbox{ pc}$ \\
B-7 &  -0.04  & 0.26 & $Z_{\sun} = 40 \mbox{ pc}$ \\
B-8 &  -0.20  & 0.31 & inverse-square weight \\
B-9 &  -0.22  & 0.33 & inverse-cubic weight \\
B-10 & -0.22  & 0.33 & exponentially-decaying weight \\
B-11 &  +0.05  & 0.32 & no weight\\
\noalign{\smallskip}
\hline
\end{tabular}
\end{flushleft}
\end{table}

\begin{figure}
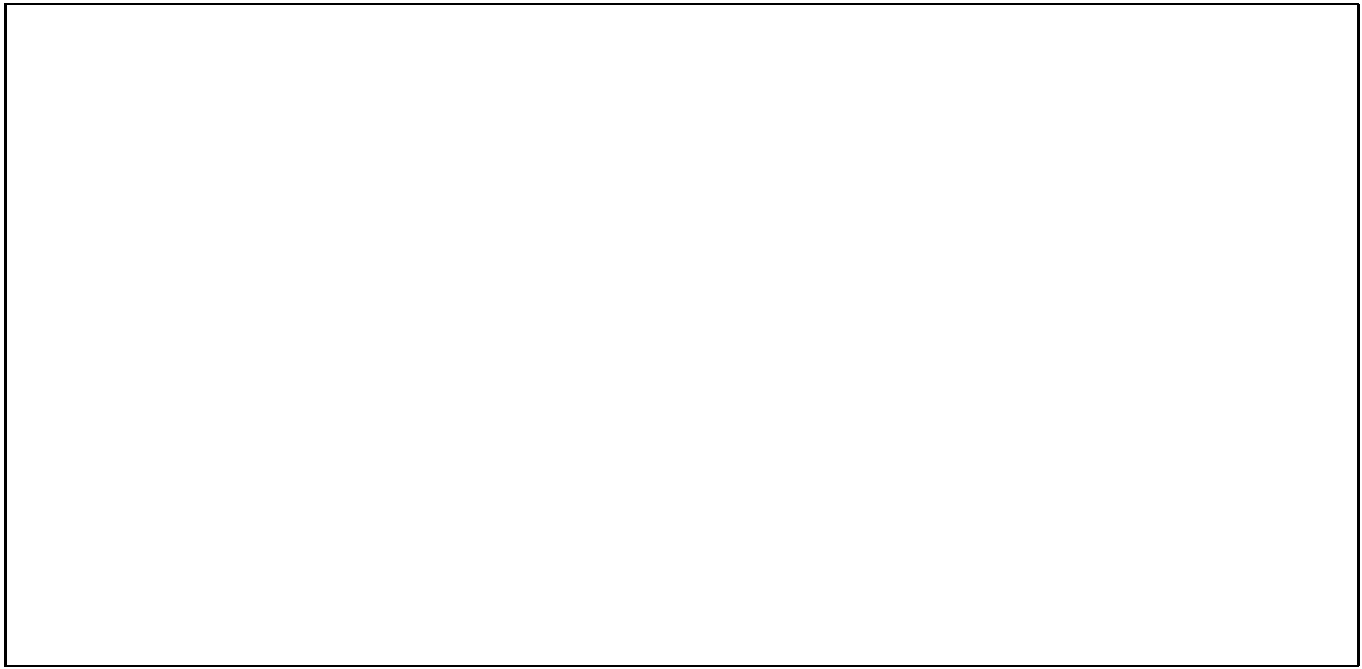

\picplace{8.8 cm}
\caption[]{Reddening residuals (in the sense $E(B-V)_{\rm model} -
E(B-V)_{\rm node}$) {\it vs.} Galactic position for our best
Model-B. Panels (a) and (b) indicate the reddening residuals {\it vs.}
Galactic longitude and latitude respectively. Panel (c) shows the
reddening residuals {\it vs.} Heliocentric distance.}
\end{figure}

\subsection{Comparisons to independent measurements of reddening}

As emphasized by J{\o}nch-S{\o}rensen (1994), a weighting \&
interpolation scheme as the one described above should give an
accurate description of interstellar reddening for fields covering a
long path through the disk or for large solid-angles, so that small
scale inhomogeneities in the distribution of dust tend to average
out. In Sect.~2.3 we already discussed the uncertainties in the
predicted values of reddening from our model for intermediate-to-high
Galactic latitudes. In this section we briefly comment on the adequacy
of our model for low Galactic latitudes, while more detailed tests are
performed through comparisons to starcounts in the Galactic plane, as
described in the following section.

J{\o}nch-S{\o}rensen \& Knude (1994) have presented a determination of
the reddening up to a distance of about 8~kpc for a field at $(l,b) =
(262^{\degr}, +4^{\degr})$ using deep uvbyH$\beta$ photometry covering
an area of $145 \, \Box '$. In Fig.~9 we show the observed upper and
lower envelope of reddening {\it vs.} Heliocentric distance as derived
from their study (their Fig.~5c), as well as the predictions from our
model. In this comparison we have used Model-B (which, in fact, gives
similar results than Model-A for distances closer than about
5~kpc). From Fig.~9 we clearly see that our method agrees quite well
with the observed {\it upper} envelope. On the other hand, a model
that does not incorporates the BH maps clearly overestimates the
reddening, showing the need of using these maps even at these low
latitudes as an extra constraint (see also
Sect.~2.5). J{\o}nch-S{\o}rensen \& Knude emphasize that the
distribution of reddening across their (small) field-of-view is
certainly not homogeneous, and that there might be at least two small
dense clouds in the field with sizes of order $5' \times 5'$. Given
these uncertainties, the agreement between our model and the upper
envelope is quite good. We also notice that the upper (observed)
envelope suggests a slope of about 0.34~mag/kpc in E(B-V) for
Heliocentric distances smaller than 2~kpc, in agreement with the
results by Knude (1979) who finds, from a multicolor study of 63
Selected Areas, that the reddening from a single average cloud is
$E(b-y)=0.03$, with 6 to 8 clouds per kpc along the line-of-sight.

\begin{figure}
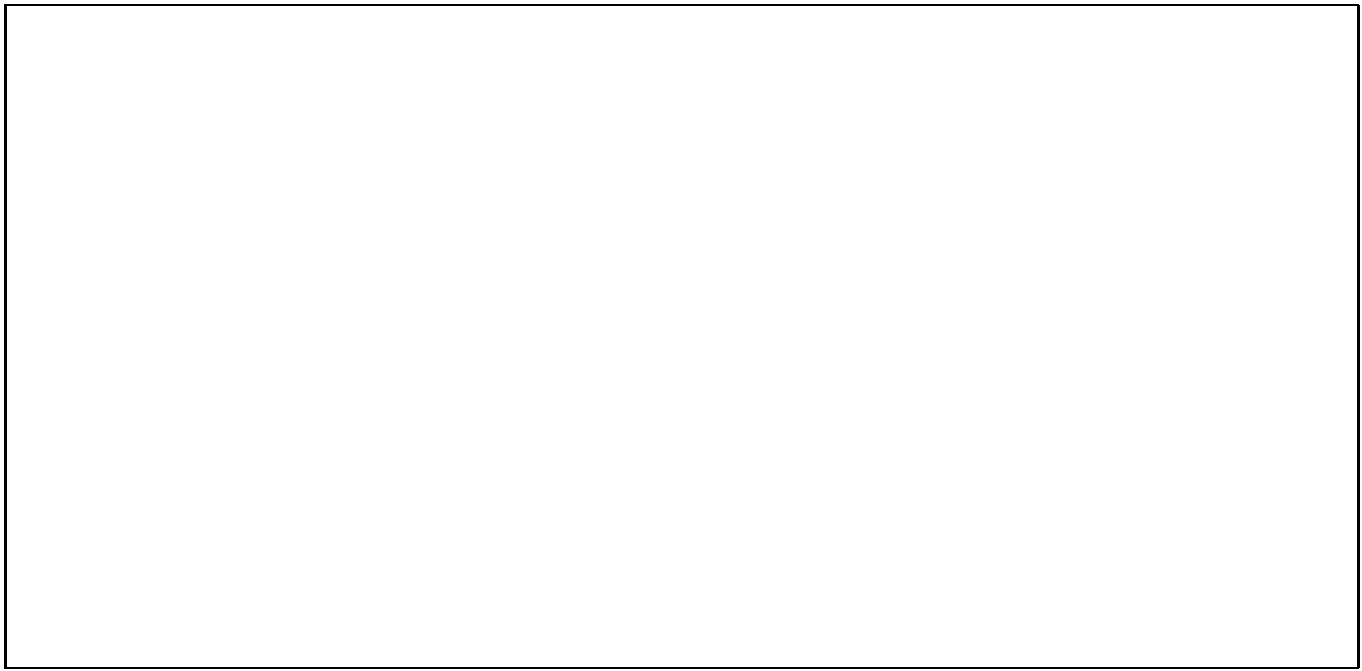

\picplace{8.8 cm}
\caption[]{Reddening {\it vs.} Heliocentric distance for a field
located at $(l,b) =(262^{\degr}, +4^{\degr}$). Solid lines indicates
the upper and lower envelope of the observed reddening as derived by
J{\o}nch-S{\o}rensen \& Knude (1994) and adopting $E(B-V) = 1.35
E(b-y)$ from Crawford (1975). The dashed line is our Model-B, while
the dot-dashed line indicates the predictions if the BH maps are not
included in the interpolation.}
\end{figure}

\section{Starcounts at low Galactic latitude: The Guide Star Catalogue}

\subsection{Selection of the Guide Star Catalogue Fields}

Perhaps one of the largest modern photographic surveys available at
low-Galactic latitudes is contained in the Guide Star Catalogue
(Lasker et al. 1990, GSC hereafter). This catalogue, covering the
whole-sky, contains calibrated photometry obtained from PDS
microdensitometer measurements of Schmidt plates taken mainly at
Siding Spring (SERC-J, UK-Schmidt Telescope, Cannon 1984) and Mount
Palomar (Quick-V, Oschin-Schmidt, Minkowski \& Abell 1963, Reid et
al. 1991).

We have chosen a number of fields from the GSC to test our reddening
model. The predictions from a starcount model (M\'endez \& van Altena
1996, M\'endez et al. 1996) are compared to actual star counts toward
these selected low-Galactic latitude regions having different amounts
of reddening. These comparisons allow us to test the extent to which
the assumptions embedded in the reddening model are appropriate, and
provide also an indication of the correctness of our representation of
the stellar populations contributing to the counts.

For our comparisons we have selected fields with plates from the {\it
supplemental survey plate collection}. These plates are not part of
the standard Palomar or SERC surveys, but were included for the
purpose of providing uniform sky coverage. Among these, we have
selected the short-exposure V plates (designated as XV on Table III of
Lasker et al. 1990) taken with the UK-Schmidt Telescope. These plates
have superior image quality which, along with their short-exposure
time (240~sec), greatly reduce confusion and classification problems,
especially severe at low-Galactic latitudes (Bernacca et al. 1992,
Lattanzi 1994 private communication). We have also looked at two other
fields from the northern Quick-V (QV) survey for completeness, but in
this case we have restricted our selection to only those fields with
plates of the type Pal-V4 which match the photometric response
function of the XV plates. The full collection of XV plates might
prove to be quite valuable for Galactic structure studies at
low-latitudes but it has, of course, limited sky coverage. The
complete sample of XV plates with $|b|< 10^{\degr}$ available from the
GSC is shown in Fig.~10, along with an indication of the fields
selected in this study.

\begin{figure}
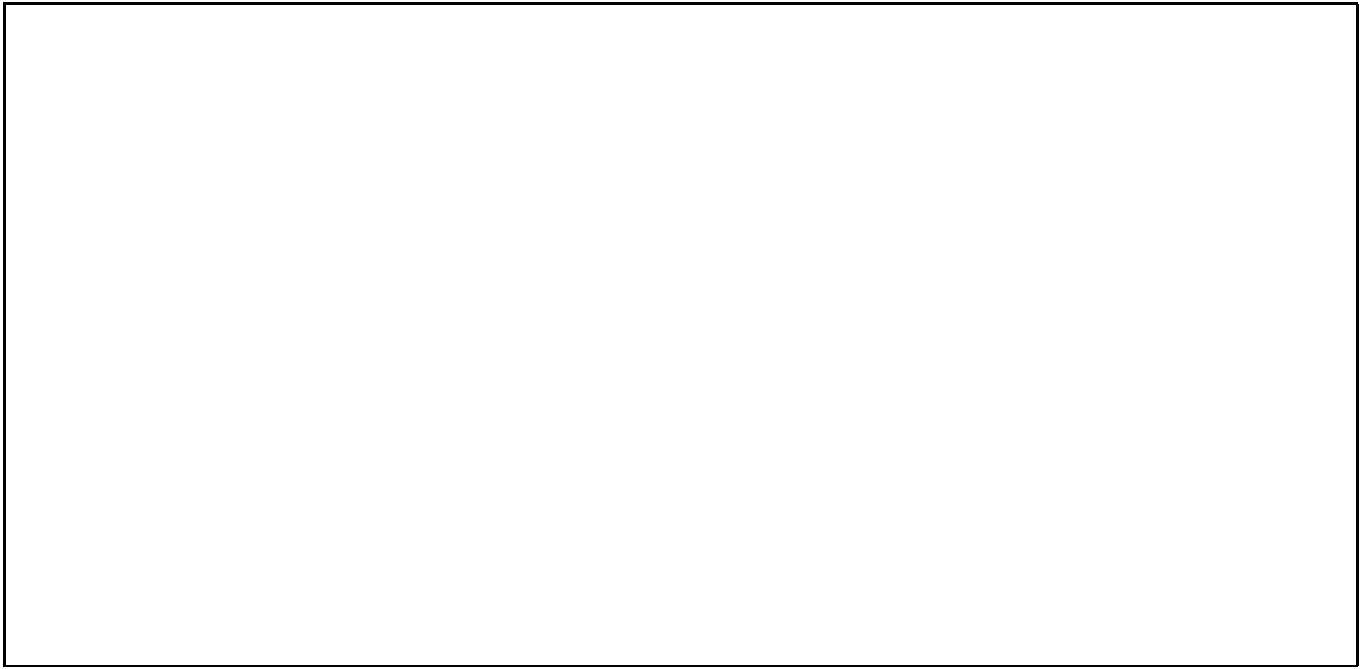

\picplace{8.8 cm}
\caption[]{Distribution on the Galactic plane of the GSC supplemental
plate collection from the UK-Schmidt Telescope (open squares, not to
scale). The fields selected for this study are marked with an X within
the open square. Two other plates, from the Palomar Quick-V survey,
have been selected as well, these are indicated as filled squares.}
\end{figure}

Six regions with centers at $|b| < 10^{\degr}$ have been selected from
the GSC to compare with our model predictions (Fig.~10). In all cases
only the $3^{\degr} \times 3^{\degr}$ central area from each plate has
been extracted in order to minimize systematic as well as random
errors in the photographic photometry which are present in regions far
from the plate's photometric sequence, usually located in the central
area of the plates (Ratnatunga 1990, Bernacca et al. 1992). The
selected solid angle is a compromise value that provides sufficiently
good number statistics (typically a few thousand stars), and small
photometric errors (typically smaller than $0.3 \mbox{ mag}$ in V for
the selected regions). These six selected regions have varying amounts
of (predicted) reddening and, therefore, provide a good test case in a
wide variety of conditions with regards to both the amount and
distribution of the reddening material along the
line-of-sight. Figure~11 shows the reddening {\it vs.} Heliocentric
distance for the selected fields, using the extrapolation method of
Model-A (which, as discussed on Sect. 2.7, gives nearly the same
results as for Model-B for distances closer than about 5~kpc). We have
also compared our reddening model to the predictions of the extinction
model by Arenou et al. (1992). Arenou's values (from his Eq. (5) and
(5bis), and the parameters given in the Appendix of their paper) are
also shown in Fig.~11. We should emphasize that, as a consequence of
our local, rather than global and analytic, treatment of the
reddening, the reddening values derived from our model, and plotted in
Fig.~11, do not reflect, necessarily, the overall value of $E(B-V)$
for the whole $3^{\degr} \times 3^{\degr}$ starcount region, but only
the values for the {\it field center} as given in Table
2. Nevertheless, Fig.~11 gives an indication of the the nature and
amount of error in the reddening estimation that one might expect from
these type of models.

\begin{figure}
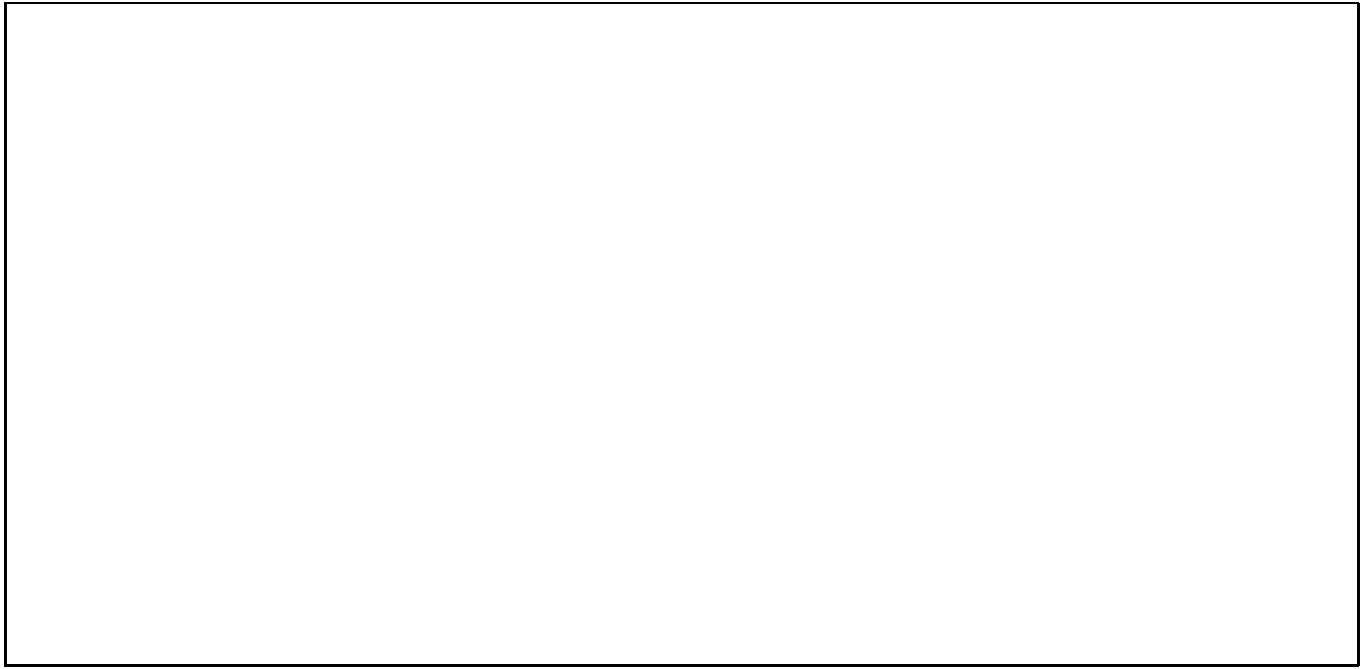

\picplace{8.8 cm}
\caption[]{Predicted reddening values for the selected
regions. Regions are numbered according to the ID given in Table
2. The solid line indicates our model predictions, while the dotted
line indicates the prediction from the extinction model by Arenou et
al. (1992) scaled to $E(B-V)$ using $R_{\rm v}$= 3.2.}
\end{figure}

Starcount data for our chosen fields have been obtained from the GSC,
version 1.2, available on the WWW
(http://www-gsss.stsci.edu/gsc12/gsc12.html), for magnitudes brighter
than $V \approx 15$. To avoid extrapolated photometric values, we have
restricted our comparisons to the range $V>8$, compatible with the
brightest stars in the photometric sequences. Only objects classified
as stars were included in the counts. Of course, at these bright
magnitudes and low-Galactic latitudes, compact galaxies misclassified
as stars do not contribute to the starcounts. On the other hand, our
comparisons will be restricted to the magnitude range where the counts
are complete, this is typically more than one magnitude {\it brighter}
than the magnitude limit of the plates. Therefore, misclassification
due to poor signal-to-noise, close blends, or plate defects is likely
to be very small (Jenkner et al. 1990).

Table 3 summarizes the properties of the selected GSC1.2 fields. The
first four columns indicate a running identification number, the GSC
source for the field, and the Galactic longitude and latitude for the
center of the field-of-view. The fifth column ($E(B-V)_{1}$)
indicates the predicted reddening in $(B-V)$ at a Heliocentric
distance of 1 kpc, as computed from our reddening model, the sixth
column ($V_{lim}$) indicates the completeness limit for the GSC
starcounts, and the seventh column indicates the observed
(error-convolved) number of stars from the GSC1.2 in the magnitude
range $8 \le V< V_{lim}$. Finally, the last column indicates a generic
name for the selected regions, with a reference to optical and/or
submillimetric studies from where the denomination has been extracted.

\begin{table*}
\caption[]{General properties of our selected GSC1.2 fields.}
\begin{flushleft}
\begin{tabular}{cccccccc}
\noalign{\smallskip}
\hline
\noalign{\smallskip}
Region ID & Source &l & b & $E(B-V)_{1}$ & $V_{lim}$ & $N_{obs}$ & Comments\\
& & deg & deg & mag & mag & [$8 \le V < V_{lim}$] & \\
\noalign{\smallskip}
\hline
\noalign{\smallskip}
1  & XV807 & 27.6 &  -0.4 & 0.66 & 12.2 &  1\,072 & Aquila Rift$^{\rm
a}$ \\
2 & QV-N338 & 67.7 &  -0.1 & 0.37 & 12.2 &  1\,771 & Cygnus Rift$^{\rm
a,b}$ \\
3 & QV-N158 & 160.8 &   5.9 & 0.43 & 13.7 & 3\,449 &
Auriga-Gemini$^{\rm b}$ \\
4  & XV168 & 283.5 &   2.2 & 0.16 & 12.2 & 2\,392 & Vela-Carina$^{\rm a,c}$ \\
5  & XV127 & 284.2 &  -3.4 & 0.13 & 12.2 & 3\,023 & Vela-Carina$^{\rm a,c}$ \\
6  & XV455 & 358.6 &   0.3 & 0.13 & 12.2$^{\rm d}$ & 1\,020 & Scorpio $^{\rm a}$ \\
\noalign{\smallskip}
\hline
\end{tabular}
\end{flushleft}
\begin{list}{}{}
\item [$^{\rm a}$] From Dame \& Thaddeus (1985).
\item [$^{\rm b}$] From Deutschman et al. (1976).
\item [$^{\rm c}$] From Dame et al. (1987).
\item [$^{\rm d}$] The completeness magnitude limit for this Region is
undefined, see Sect 3.2.
\end{list}
\end{table*}

We should emphasize that the photometry from the GSC is in the
``natural system'' of each filter-emulsion combination. Therefore, for
the XV plates (emulsion IIaD, filter GG495), V refers only to an
approximation to Johnson's V, and the same is true for the Pal-V4
plates (with the same emulsion-filter combination). Therefore, in what
follows, whenever we mention V photometry it must be understood that
we are referring to the ``natural'' passbands of these plates. We
should also emphasize that all our model predictions (see Sect 3.2)
have been converted to the photometric systems of the plates, by using
the following transformation (Lasker et al. 1990, Table I, and Russell
et al. 1990, Eq. (1)):

\begin{equation}
V_{495} = V - 0.10(B-V)
\end{equation}

Notice that, in the above equation, all quantities are reddened and
extincted.

\subsection{Starcounts and the reddening model}

The starcounts model described by M\'endez \& van Altena (1996) has
been used, along with the reddening model described above, to compute
the expected counts in the selected fields. Because of the rather
large solid angle covered by each plate, and because of the discrete
nature of the reddening material, some experimentation was needed to
decide upon the best angular resolution to employ in the integration
scheme in the starcounts model. It was found that, for a resolution
higher than $0.5^{\degr}$ in both RA and DEC, the predicted counts
from the model varied by less than 2\%. Therefore, we adopted this
angular resolution. It must be emphasized that a finer angular
resolution could easily be performed with the model, but then the
required CPU time becomes increasingly large, making these
computations impractical (e.g., already, at $0.5^{\degr}$ resolution,
the model has to be evaluated at 36 positions within each Region). Our
adopted angular resolution {\it oversamples} the mean angular
resolution of our reddening maps (see Sect. 2.4, and Fig.~2), so that
there is no loss of information from the reddening maps when adopting
this angular resolution. Also, we have adopted Model-A for
extrapolation at larger distances although, in practice, it does not
make any difference for this application to use either Model-A or
Model-B because both models give nearly the same reddening at
distances smaller than about 5~kpc (see Sect.3.1), while most of the
starcounts analyzed here are contributed by objects closer than about
2~kpc (see Sect.~3.3). The adoption of Model-A over Model-B has been
guided by the greater simplicity of the former, and a faster
implementation of the algorithm (compare Eq.~(6) for Model-A with
Eq.~(7) for Model-B), although for general applications Model-B would
lead to better predictions than Model-A for reasons that were
explained in Sect.~2.5.2. The Galactic structure parameters employed
in our computations are those described in M\'endez \& van Altena
(1996, Table 1), unless otherwise specified.

Starcounts {\it vs.} magnitude for the selected regions are shown in
Fig.~12. The predicted counts are for the standard parameters of the
reddening procedure as outlined in the previous sections. These
comparisons indicate a very good agreement between the predicted and
observed starcounts in the magnitude range $8 \le V < 14.5$,
especially when considering the model predictions for the case of
no-reddening. Table~4 indicates the observed starcounts within the
magnitude range where the counts are complete (i.e., $8 \le V <
V_{lim}$, from Table~3) and the predicted starcounts by the model
without (Run~1) and with (Run~2) reddening, in the same magnitude
range. The last two columns in Table~4 indicate the mean fractional
difference between the observed and model starcounts, as well as the
mean of the absolute dispersion, both with their standard
deviation.\footnote{In Tables~4 and~5 the values quoted in the last
two columns {\it exclude} Region~6 for reasons that will become
apparent afterwards.} It can be seen that, despite a reduction factor
of up to 15 in the predicted counts from a model without reddening to
a model including reddening, the (absolute) agreement is around 20\%,
with a mean value of 11\% (except in the case of Region~6 at $(l,b) =
(358.6^{\degr}, +0.3^{\degr})$ which will be discussed later).

\begin{figure}
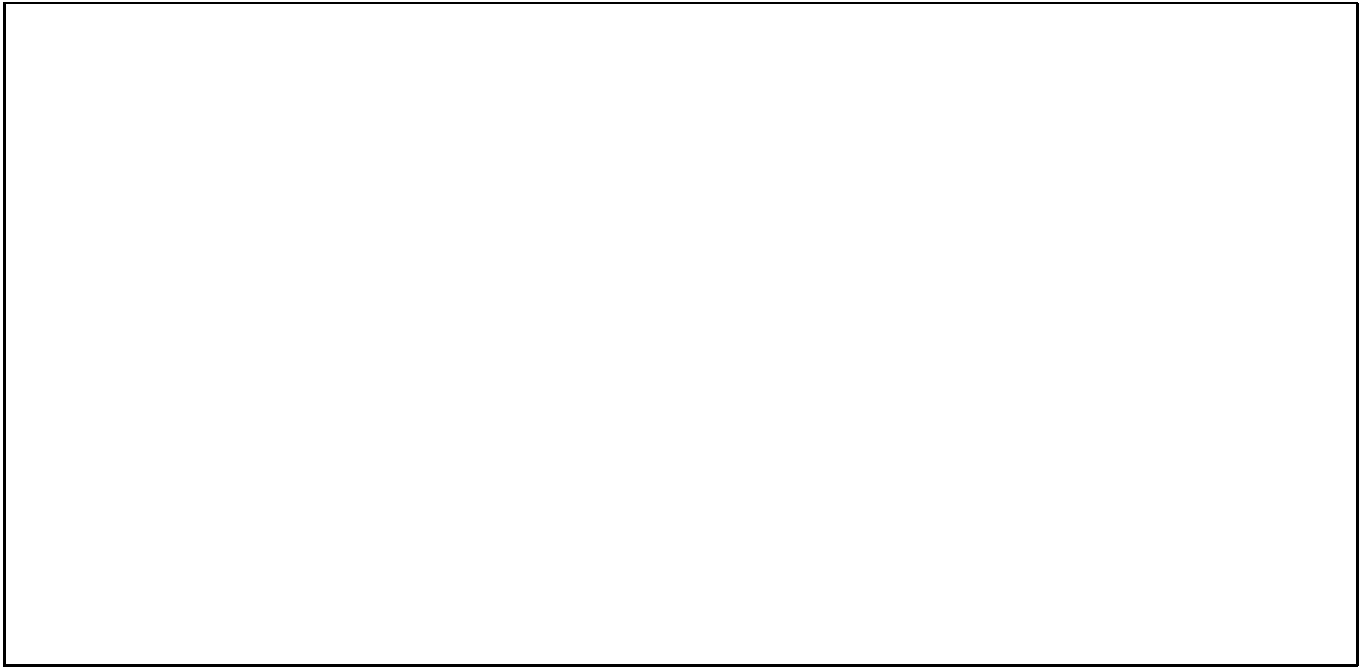

\picplace{8.8 cm}
\caption[]{Observed (dots with Poisson error bars) and predicted
(lines) starcounts in our selected regions. The solid line indicates
the predicted counts for no-reddening (Run~1), while the dotted line
indicates the predictions using the model described in Sect.~2
(Run~2).}
\end{figure}

Starcounts in our selected fields are somewhat sensitive to the value
of $R_{\rm v}$. It is well known (Guti\'erez-Moreno \& Moreno 1975)
that $R_{\rm v}$ is as function of, both, the spectral energy
distribution of the stars considered, and the amount of reddening
itself. Figure~13 shows the results of adopting the (mean)
relationship for $R_{\rm v}$ given by Schmidt-Kaler (1982):

\begin{equation}
R_{\rm v} = 3.30 + 0.28(B-V)_{\rm o} + 0.04 E(B-V)
\end{equation}

\begin{figure}
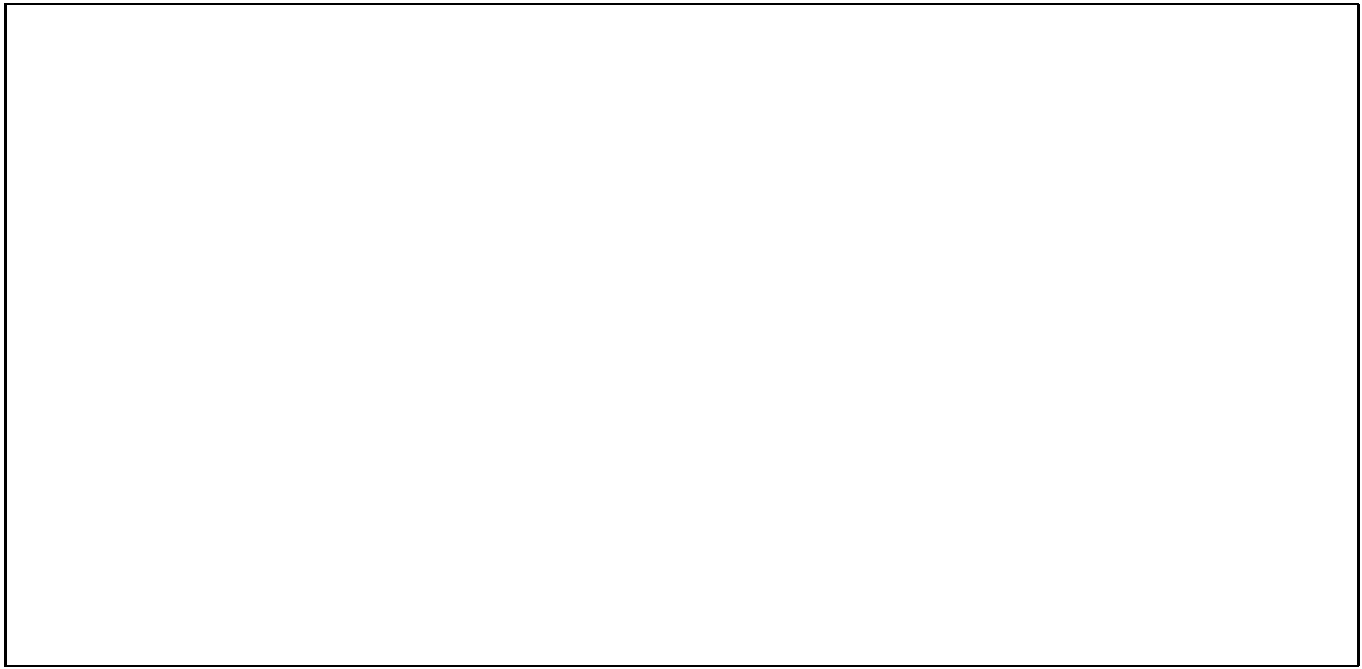

\picplace{8.8 cm}
\caption[]{Same as Fig. 12, except that $R_{\rm v}$ is as given by
Eq. (17, solid line). A change on $R_{\rm v}$ of $\pm 0.2$ is
indicated with the dashed lines, with the upper predicted counts
having the smaller value of $R_{\rm v}$.}
\end{figure}

By comparing Figs. 12 and 13 (see also Table 4), it is evident that we
obtain a better fit to the observed counts by using $R_{\rm v}$ as per
Eq. (17), rather than a constant value. The dependency on $R_{\rm v}$
is not surprising, in view of the large amount of reddening present in
some of our fields. Indeed, it is clear from Fig. 13 that the
dependency on $R_{\rm v}$ is larger for those fields with larger
overall reddening (Table 3 and Fig. 11). Schmidt-Kaler cites regional
variation of $R_{\rm v}$ amounting to $\pm 0.2$. In Fig.~13 (see also
Table~4) we show the effect of changing $R_{\rm v}$ within these
margins. The effect of these changes is rather small, and it suggests
that a better overall model is indeed obtained with an $R_{\rm v}$
that is given by Eq. (17). Therefore, we have adopted this value of
$R_{\rm v}$ for {\it all} our subsequent model runs.

At this point it is interesting to mention that the predictions of a
model similar to that of Run~3 (which uses the $R_{\rm v}$ given by
Eq.~(17)), but using the extrapolation Model-B, leads to a mean
difference in the counts for all six regions of only 1.7\% with
respect to Run~3, while the mean absolute difference is 4\%. This
shows that the use of Model-A, for this particular application, is
guaranteed. However, comparisons to fainter starcounts would likely
benefit from the use of Model-B instead, for the reasons presented in
Sect.~2.5.2.

\begin{table*}
\caption[]{Predicted starcounts as a function of parameters in the
reddening model.}
\begin{flushleft}
\begin{tabular}{cccccccccc}
\noalign{\smallskip}
\hline
\noalign{\smallskip}
Run & Reg. 1 & Reg. 2 & Reg. 3 & Reg. 4 & Reg. 5 & Reg. 6 &
$<\frac{\Delta N}{N_{\rm Obs}}>^{\rm a,b}$ & $<\frac{| \Delta N
|}{N_{\rm Obs}}>^{\rm b}$ & Comments$^{\rm c}$\\
\noalign{\smallskip}
\hline
\noalign{\smallskip}
Obs. & 1\,072 & 1\,771 & 3\,449 & 2\,392 & 3\,023 & 1\,020 & -- & -- & Measured values\\
1 & 22\,164 & 12\,530 & 7\,925 & 7\,231 & 6\,099 & 30\,134 & $6.0 \pm
7.9$ & $6.0 \pm 7.9$ & No reddening\\
2 &  1\,459 &  2\,181 & 2\,782 & 2\,989 & 2\,659 &  3\,162 & $0.11 \pm
0.25$ & $0.23 \pm 0.09$ & Std. params., $R_{\rm V}=3.2$\\
3 &  1\,296 &  1\,977 & 2\,489 & 2\,780 & 2\,510 &  2\,888 & $0.01 \pm
0.22$ & $0.19 \pm 0.06$ & $R_{\rm V}$ as in Eq. (17)\\
4 &  1\,164 &  1\,832 & 2\,332 & 2\,658 & 2\,425 &  2\,651 & $-0.06 \pm
0.19$ & $0.15 \pm 0.11$ & $R_{\rm V}$ as in Eq. (17) +0.2\\
5 &  1\,437 &  2\,142 & 2\,658 & 2\,911 & 2\,600 &  3\,164 & $0.08 \pm
0.25$ & $0.23 \pm 0.07$ & $R_{\rm V}$ as in Eq. (17) -0.2\\
6 &  1\,234 &  1\,814 & 2\,215 & 2\,745 & 2\,509 &  2\,644 & $-0.04 \pm
0.22$ & $0.17 \pm 0.12$ & $\delta A_{\rm v} =1.5$ mag/kpc\\
7 &  1\,404 &  2\,147 & 2\,335 & 2\,856 & 2\,503 &  2\,991 & $0.04 \pm
0.28$ & $0.24 \pm 0.07$ & $h_{\rm red}= 50$~pc\\
8 &  1\,234 &  1\,882 & 2\,757 & 2\,744 & 2\,539 &  2\,764 & $0.00 \pm
0.17$ & $0.14 \pm 0.05$ & $h_{\rm red}= 200$~pc\\
9 &     770 &  1\,442 & 2\,851 & 2\,455 & 2\,279 &  2\,511 & $-0.17 \pm
0.12$ & $0.18 \pm 0.10$ & Arenou et al. (1992) model\\
\noalign{\smallskip}
\hline
\end{tabular}
\end{flushleft}
\begin{list}{}{}
\item [$^{\rm a}$] $\Delta N$ is in the sense $N_{\rm Model}-N_{\rm
Obs}$.
\item [$^{\rm b}$] These columns exclude Region 6, see text for
explanation.
\item [$^{\rm c}$] The meaning of the different runs is discussed in
the text. Here we give only a brief description.
\end{list}
\end{table*}

We have also investigated the effects on the predicted starcounts of
changing the differential absorption and the scale-height of the
reddening material, since these parameters were not well constrained
by the tests described in Sect.~2.6. Figure~14 shows the results of
our runs with $\delta A_{\rm v}=1.5$~mag/kpc (Run~6 in Table 4), as
well as the effect of changing $h_{\rm red}$ from 110~pc (Run~3 in
Table 4) to a lower value of 50~pc (Run~7 in Table 4), and to an upper
value of 200~pc (Run~8 in Table~4). From the last two columns on
Table~4 it is apparent that the change of $\delta A_{\rm v}$ has a
minor impact on the predicted counts. However, there is a tendency of
the model with the larger value of $\delta A_{\rm v}$ to underestimate
the counts, specially at the fainter magnitudes. This is a natural
consequence of the excessive differential reddening implied by this
value of $\delta A_{\rm v}$, suggesting that the value of 0.5~mag/kpc
found in Sect.~2.6.1 is closer to the proper value. The lack of
sensitivity of the predicted counts to this parameter is
understandable because this differential reddening is only applied
beyond the last node point in the reddening maps, which extend to
distances beyond 2~kpc, while the mean distance of stars contributing
to the observed counts is typically less than 2~kpc, reinforcing thus
the adequacy of using Model-A for reddening extrapolation.

With regards to the scale-height of the reddening material, a higher
value for $h_{\rm red}$ seems to provide a better overall solution in,
both, the mean as well as the absolute residuals. As we shall see
though (Sects.~3.3 and~3.4), there are other parameters that also have
a large impact on the starcounts. Therefore, in what follows, we will
still adopt our standard values of $\delta A_{\rm v}=0.5$~mag/kpc and
$h_{\rm red}=110$~pc.

\begin{figure}
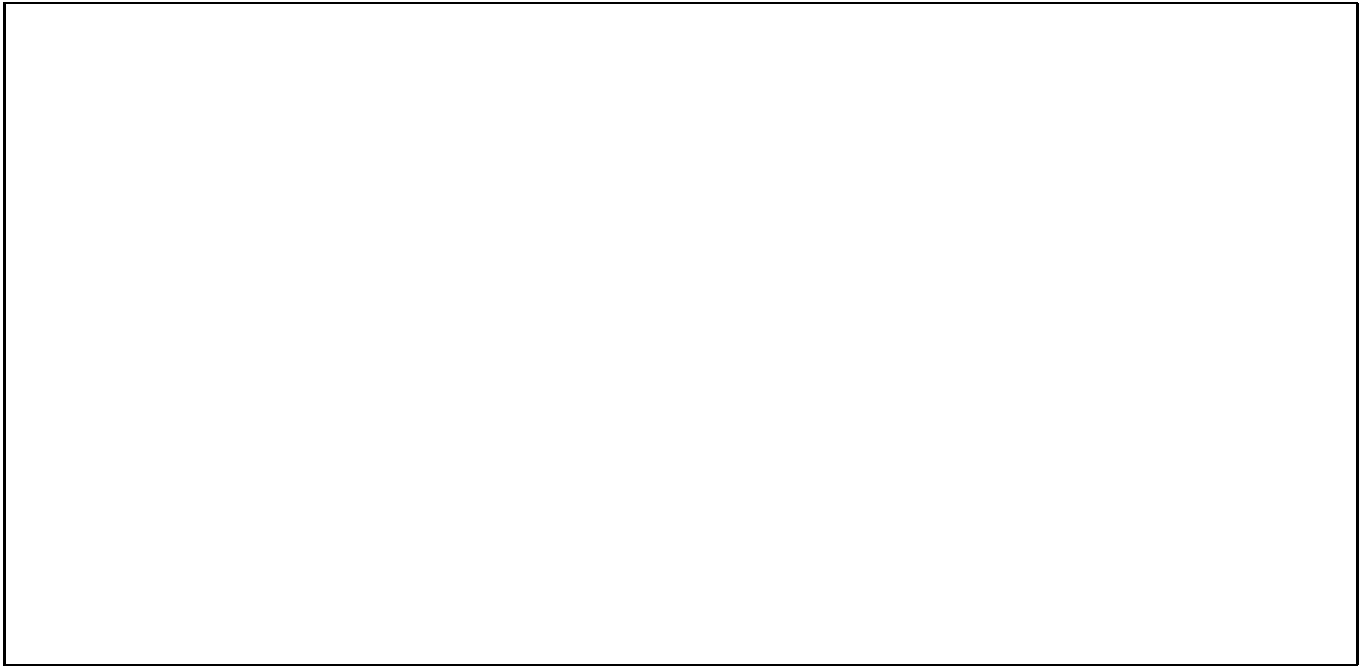

\picplace{8.8 cm}
\caption[]{Same as Fig. 13, except that $\delta A_{\rm v}$ has been
increased to 1.5~mag/kpc (solid line). The dashed line indicates the
run with a small $h_{\rm red}$ equal to 50 pc, while the dot-dashed
line indicates the run with a large $h_{\rm red}$ equal to 200 pc,
both having $\delta A_{\rm v}=0.5$~mag/kpc, as in Figs. 12 and 13.}
\end{figure}

Our starcounts model also includes an option for the extinction model
by Arenou et al. (1992). This model was developed to predict
magnitudes and colors for stars in the Hipparcos Input Catalogue,
which in many cases lacked photoelectric photometry. The results of
these computations, along with the results from Run~3, are indicated
in Fig.~15 \& Table~5. There is a very good correspondence between the
starcounts predicted using our reddening model and Arenou's et
al. model, specially when considering the rather large differences in
the predicted reddening values between these two models shown in
Fig. 11. However, it is also apparent from Fig.~15 (see also Table 4,
Run~9) that Arenou's model predicts systematically {\it fewer} stars
than observed, although the absolute dispersion with respect to the
observed value is similar to that of our best runs (e.g., Runs~3
and~8), as indicated in the last two columns of Table~4. The
discrepancies in the predicted starcounts between these two reddening
models may be a consequence of the different approaches adopted: while
Arenou's et al. model accounts for the large-scale properties of the
reddening material, our model incorporates more accurately the
intrinsic clumpiness of the absorbing material, giving therefore
better results {\it on the mean} at low Galactic latitudes, even for
very high-reddening regions. As a general statement we could say that
the {\it absolute} deviation in the predicted starcounts from the
observed values is somewhat smaller when using the model by Arenou et
al. than our model. This is in agreement with the fact that Arenou et
al. (1992) quote a {\it mean accuracy} for their model of about 40\%,
which for a reddening at 1~kpc of, e.g., 0.4~mag, leads to an
uncertainty in the reddening of 0.16~mag, smaller almost by a factor
of 1.5 than the estimated uncertainty of 0.23~mag for our model
reddening values (see Sect. 2.6 and Fig. 7). However, the deviations
between the starcounts computed with the model by Arenou et al. and
the observed counts seem to be systematic and, therefore, even though
their model leads to more precise starcounts, our model seems to lead
to more accurate starcounts, in the classical statistical distinction
between these two terms.

We must emphasize that Arenou et al. (1992) have adopted the mean
$R_{\rm v}$ as given in Eq. (17) above. Since Arenou's model is an
extinction model, rather than a reddening model, we have to adopt an
iterative procedure to solve for $E(B-V)$ (and thus $({\rm
B-V})_{\rm o}$) from their model $A_{\rm v}$ values and from the
observed (B-V). This was needed in order to compute starcounts
with our model, and it is a consequence of the dependence of $R_{\rm
v}$ on both $(B-V)_{\rm o}$ and $E(B-V)$. The iterative
procedure converged quickly, in only two or three iterations.

\begin{figure}
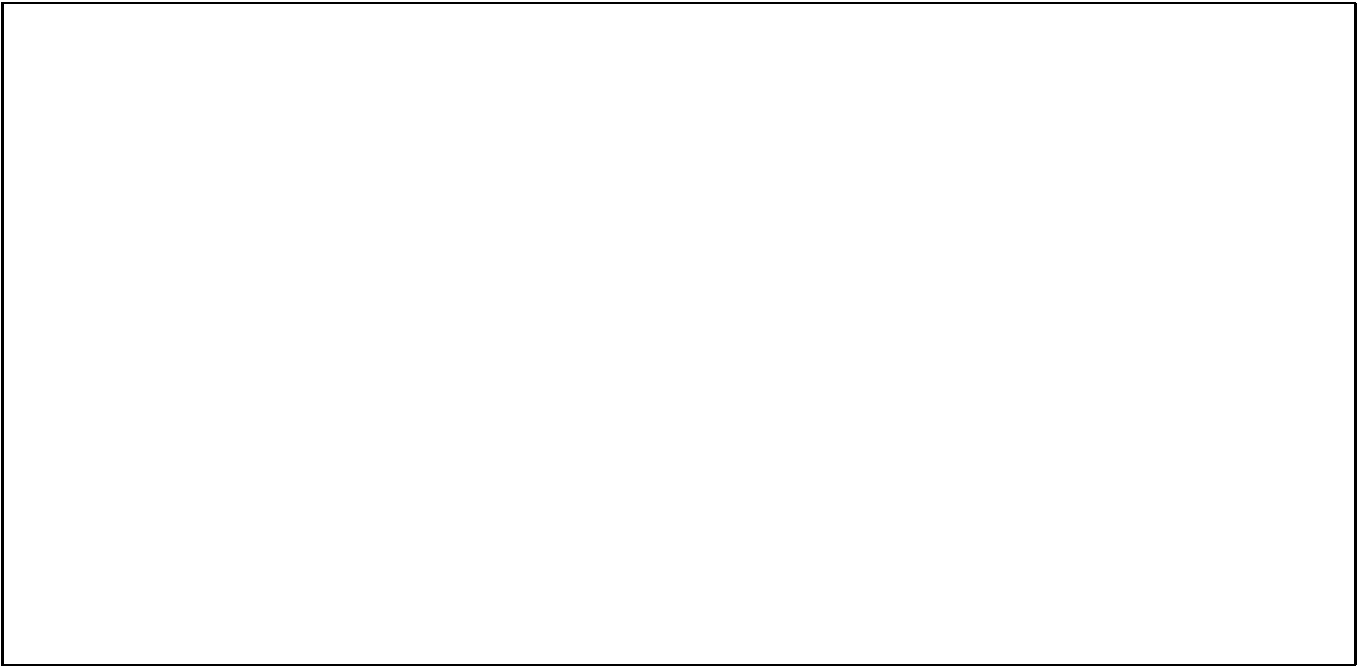

\picplace{8.8 cm}
\caption[]{Same as Fig. 13, except that the extinction model by Arenou
et al. (1992) is used (solid line, Run~9). For comparison, we show the
results from Run~3 as a dashed line.}
\end{figure}

The discrepancy seen between the model predictions and Region~6 at
$(l,b) = (358.6^{\degr}, +0.3^{\degr})$ (Figs. 13 -- 15) is puzzling:
on one hand the predicted reddening at 1~kpc is smaller than that of
Region~1 at $(l,b) = (27.6^{\degr}, -0.4^{\degr})$ (Table 3), yet the
observed number of stars in the latter is larger than that of the
former, contrary to expectation. On the other hand, Region~6 should
exhibit a higher stellar density than that of Region~1 (before any
reddening correction) because of the exponentially decaying stellar
density profile away from the Galactic center (this is indeed seen in
the model excluding reddening on Fig. 12 and Table 4, Run~1). These two
facts combined, point to a problem with the particular GSC plate in
Region~6, rather than to a deficiency of the model. It is possible
that due to the very high degree of crowding toward this line of
sight, only a subset of all images were included in the GSC in order
to maintain a good magnitude coverage (one of the GSC requirements was
to provide a uniform stellar density over the sky, rather than a
uniform magnitude limit, so that at lower Galactic latitudes the GSC
becomes incomplete at brighter magnitudes than for higher Galactic
latitudes (Jenkner et al. 1990). An investigation of this particular
plate is currently being carried out to determine the actual nature of
the discrepancy seen in Figs. 13 -- 15.

\subsection{Starcounts and Galactic Structure}

From Eqs. (4),(7), and (13) it is clear that the value of $Z_{\sun}$
will have an impact on the predicted reddening. Indeed, this parameter
will have an important effect, even if there is no reddeding, as it
controls the viewing aspect when looking at the Galactic plane at
positions slightly above or below it. Therefore, its effect is
twofold: it changes the predicted value of $E(B-V)$, but it also
changes the density contribution of the different types of stars that
make up the starcounts model. Table~5 indicates the predicted
reddening for a model with the same parameters as that for Run~3 in
Table~4, but for values of $Z_{\sun}=+40 \mbox{ pc}$ and $Z_{\sun}=-10
\mbox{ pc}$ (instead of the standard value $Z_{\sun}=+7 \mbox{ pc}$
adopted in Runs~1 to~9 of Table~4). Figure~16 show the effects on the
predicted starcounts. From the last column on Table~5 it is evident
that we obtain a more precise fit for a value of $Z_{\sun}=+40 \mbox{
pc}$, while a value of $Z_{\sun}=-10 \mbox{ pc}$ is clearly
inadequate.  The latter value has been recently suggested by Binney et
al. (1996) who have found, from a dust-corrected near-infrared
COBE/DIRBE surface brightness map of the inner Galaxy ($|l| \le
30^{\degr}$, $|b| \le 15^{\degr}$), that the Sun should lie $14 \pm
4$~pc {\it below} their (photometric-model) symmetry plane. This
solution seems incompatible with the GSC low-latitude optical
starcounts presented here. Indeed, starcount studies at the Galactic
poles indicate that the Sun is {\it above} the Galactic plane (as
defined by the distribution of neutral Hydrogen), e.g., Humphreys \&
Larsen (1995) find $Z_{\sun}= 20.5 \pm 3.5 \mbox{ pc}$ from a
comparison of the asymmetry in the starcounts towards the North and
South Galactic poles.

\begin{figure}
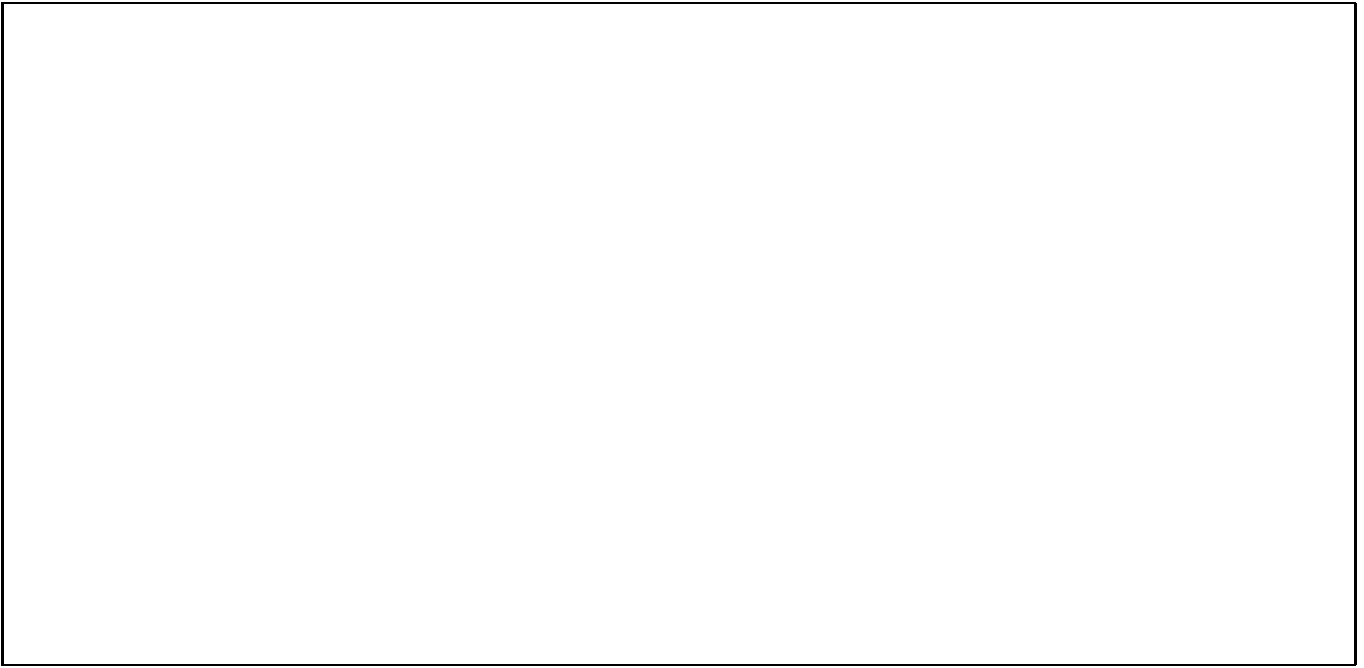

\picplace{8.8 cm}
\caption[]{Same as Fig. 13, except that we have used $Z_{\sun}=+7
\mbox{ pc}$ (solid line, Run 3) $Z_{\sun}=+40 \mbox{ pc}$ (dashed
line, Run 10), and $Z_{\sun}=-10 \mbox{ pc}$ (dot-dashed line, Run
11).}
\end{figure}

We have also varied the adopted value of $R_{\sun}=8.5$~kpc within a
wide margin, between 7~kpc and 10~kpc. The effect of these
self-similar Galaxy models on the starcounts is minimal, as it can be
seen in Table~5, with no salient features in the shape of the
predicted starcounts. This is a consequence of the relatively small
distance range covered by these bright starcounts, limited typically
to Heliocentric distances smaller than about 2~kpc.

The structural parameters discussed previously affect {\it all} the
stellar populations simultaneously. However, there are still a few
population-dependent parameters in the starcount model that have
an impact on the starcounts: these are the population scale-heights
and scale-lengths.

Figure~17 shows the effect of varying the exponential scale-length for
the disk ($H_{\rm Disk}$) between 2.5~kpc and 4.5~kpc. The effect on
the counts is highly dependent on projection effects, and thus
Regions~1 and~6 are the most affected by changes in this parameter,
along with Region~3 at the fainter magnitudes, all these Regions being
close to the center-anticenter direction. A {\it longer} $H_{\rm
Disk}$ of about 3.5 to 4.5~kpc seems to reproduce better the counts,
rather than the short-scale of $2.3 \pm 0.1$~kpc advocated recently by
Ruphy et al. (1996) from star and color counts from the DENIS
project. Since all these determinations {\it are} model-dependent, it
is possible that other Galactic parameters have come into play in the
determination by Ruphy et al., since their fields were located at $l=
217^{\degr}$ and $l= 239^{\degr}$, close to our Regions~4 and~5 were
we see (Fig.~17) a reduced dependency of the predicted counts on
$H_{\rm Disk}$. This highlights the importance of studying {\it
several} directions simultaneously to best constrain the structural
properties of the Galaxy from the vast parameter space available in
the models. On the other hand, our results agree with the
determination by Ng et al. (1995), who obtained a value of $H_{\rm
Disk}$ in the range 3.5 -- 4.5~kpc from an analysis of starcounts
towards the Galactic Bulge (see Sect. 3.4).

\begin{table*}
\caption[]{Predicted starcounts as a function of Galactic Structure
parameters in the starcounts model.}
\begin{flushleft}
\begin{tabular}{cccccccccc}
\noalign{\smallskip}
\hline
\noalign{\smallskip}
Run & Reg. 1 & Reg. 2 & Reg. 3 & Reg. 4 & Reg. 5 & Reg. 6 &
$<\frac{\Delta N}{N_{\rm Obs}}>^{\rm a,b}$ & $<\frac{| \Delta N
|}{N_{\rm Obs}}>^{\rm b}$ & Comments$^{\rm c}$\\ \\
\noalign{\smallskip}
\hline
\noalign{\smallskip}
Obs. &  1\,072 &  1\,771 & 3\,449 & 2\,392 & 3\,023 & 1\,020 & -- & --
& Measured value\\
3 &  1\,296 &  1\,977 & 2\,489 & 2\,780 & 2\,510 &  2\,888 & $0.01 \pm
0.22$ & $0.19 \pm 0.06$ & $R_{\rm V}$ as in Eq. (17)\\
10   &  1\,081 &  1\,639 & 2\,113 & 2\,232 & 2\,951 & 2\,171 & $-0.11 \pm
0.16$ & $0.11 \pm 0.16$ & $Z_{\sun}=+40$~pc\\
11   &  1\,245 &  2\,005 & 2\,695 & 3\,125 & 2\,265 & 3\,116 & $0.03 \pm
0.25$ & $0.21 \pm 0.07$ & $Z_{\sun}= -10$ pc\\
12   &  1\,275 &  1\,961 & 2\,489 & 2\,756 & 2\,497 & 2\,846 & $0.00 \pm
0.21$ & $0.18 \pm 0.06$  & $R_{\sun}= 7$~kpc\\
13   &  1\,293 &  1\,991 & 2\,491 & 2\,797 & 2\,521 & 2\,487 & $0.01 \pm
0.22$ & $0.19 \pm 0.06$  & $R_{\sun}= 10$~kpc\\
14   &  1\,470 &  2\,050 & 2\,229 & 2\,830 & 2\,560 & 3\,540 & $0.04 \pm
0.29$ & $0.24 \pm 0.11$  & $H_{\rm Disk}=2.5$~kpc\\
15   &  1\,201 &  1\,940 & 2\,655 & 2\,753 & 2\,484 & 2\,559 & $-0.01 \pm
0.18$ & $0.16 \pm 0.05$  &  $H_{\rm Disk}=4.5$~kpc\\
16   &  1\,274 &  1\,951 & 2\,106 & 2\,590 & 2\,329 & 2\,806 & $-0.05 \pm
0.25$ & $0.20 \pm 0.12$  & $h_{\rm Disk}$ for RG \& SG$^{\rm d}$ of 150~pc\\
17   &  1\,292 &  1\,991 & 2\,725 & 2\,874 & 2\,602 & 2\,865 & $0.04 \pm
0.20$ & $0.18 \pm 0.05$ & $h_{\rm Disk}$ for RG \& SG of 350~pc\\
18   &  1\,184 &  1\,838 & 2\,035 & 2\,348 & 2\,139 & 2\,517 & $-0.12 \pm
0.22$ & $0.17 \pm 0.17$ &  Lower $h_{\rm Disk}$ for MS$^{\rm e}$ stars\\
19   &  1\,328 &  2\,032 & 2\,757 & 2\,990 & 2\,698 & 2\,986 & $0.07 \pm
0.21$ & $0.19 \pm 0.06$  & Upper $h_{\rm Disk}$ for MS stars\\ 
20   &  1\,035 &  1\,719 & 2\,557 & 2\,389 & 2\,751 & 2\,009 & $-0.08 \pm
0.10$ & $0.08 \pm 0.10$ & Best-fit solution$^{\rm f}$ \\
\noalign{\smallskip}
\hline
\end{tabular}
\end{flushleft}
\begin{list}{}{}
\item[$^{\rm a}$] $\Delta N$ is in the sense $N_{\rm Model}-N_{\rm
Obs}$. 
\item [$^{\rm b}$] These columns exclude Region 6, see text for explanation.
\item [$^{\rm c}$] The meaning of the different runs is discussed in
the text. Here we give only a brief description.
\item[$^{\rm d}$] RG stands for red-giants, SG for sub-giants.
\item[$^{\rm e}$] MS stands for main-sequence.
\item [$^{\rm f}$] From Regions~1, 2, and~4 only, with $Z_{\sun}=+27$~pc and $H_{\rm Disk}=6$~kpc,
see text.
\end{list}
\end{table*}

\begin{figure}
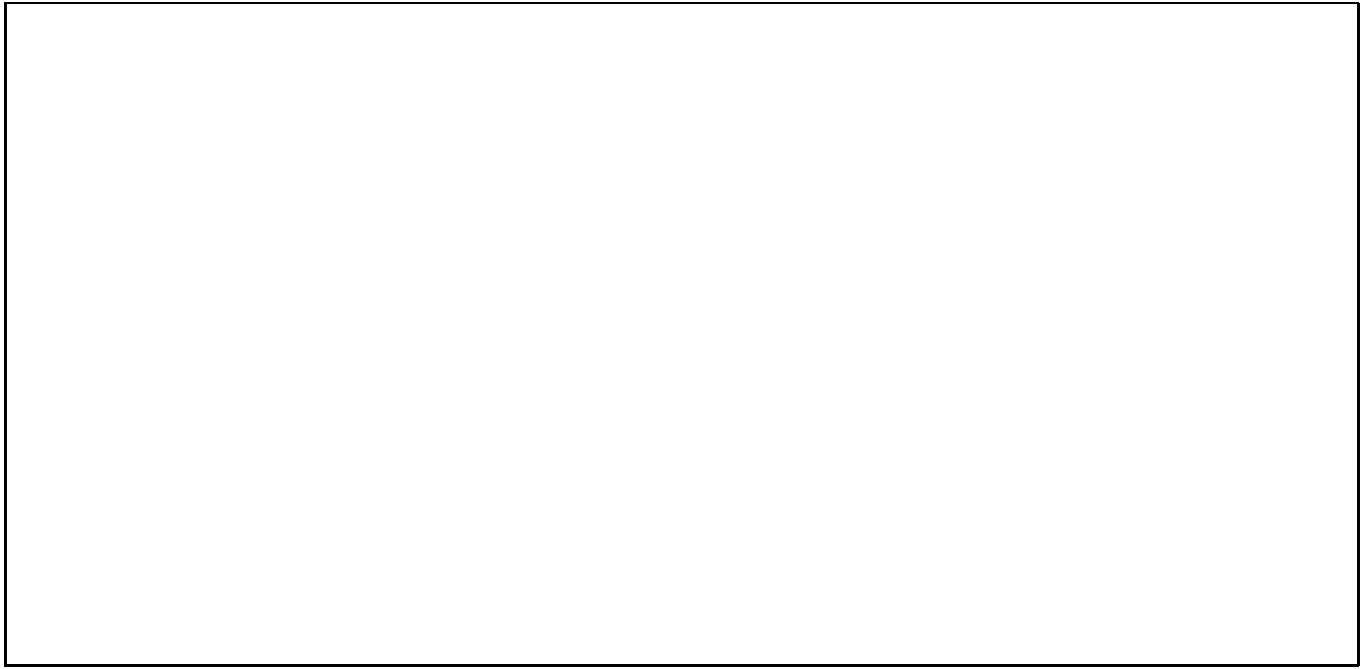

\picplace{8.8 cm}
\caption[]{Same as Fig. 13, except that we have used a scale-length
for the disk of 3.5~kpc (solid line, Run 3), 2.5~kpc (dashed line, Run
14), and 4.5~kpc (dot-dashed line, Run 15).}
\end{figure}

The scale-height for disk stars ($h_{\rm Disk}$) is parametrized as a
function of absolute visual magnitude, as described by M\'endez \& van
Altena (1996). Red-giant \& sub-giant stars (defined as $M_v \le +3.0$
and $(B-V)_o > +0.57$) are given a single $h_{\rm Disk}$ of 250~pc. We
have varied this parameter between 150~and 350~pc and found that the
effect of these changes on the predicted counts to be minimal, except
for Region~3. Region~3, with a higher Galactic latitude, is more
sensitive to this parameter as the line-of-sight includes a wider
sampling of stars in the direction perpendicular to the plane. In
fact, this effect has been used by M\'endez \& van Altena to constrain
$h_{\rm Disk}$ for disk sub-giants, obtaining a value of $250 \pm
32$~pc, in agreement with previous determinations of the scale-height
for red-giants.

As for the scale-height of main-sequence stars, we have changed this
parameter between a ``lower'' and an ``upper'' set of values allowed
by the observational uncertainties, as described by M\'endez \& van
Altena (1996). Given the distance range predicted from the model, the
luminosity range sampled by these counts encompass the range $-1.0 \le
M_v < +0.5$, i.e., it includes main-sequence stars of spectral type B5
to A0 and red-giants (luminosity class III) of spectral types G0 to
late M, located at distances of between 400~pc and 1.8~kpc. We find
that the effect of changing the scale-height of main-sequence stars
yields a systematic change in the counts, and therefore it does not
produce a good match to the overall counts. This can be clearly seen
by inspection of the last two columns on Table~5 for Runs~18 and~19.

\subsection{Putting it all together}

The purpose of the parameter-space exploration described in the
previous two sub-sections has been to indicate the possible range of
uncertainty expected from the model starcounts given our present
knowledge of the different Galactic structure and reddening-related
parameters employed in the model. It is clear from the analysis
presented above that, in principle, several combinations of parameters
can yield similar predicted starcounts and, for this reason, we have
not attempted, at this stage, a global solution. However, the numerous
trends described above do yield some constraints on parameters of the
reddening model as well as parameters of the Galactic model, and we
describe these in turn.

From the reddening model, we find that the adopted value of the
scale-height of the reddening material has an important role in the
predicted counts. On the other hand, these bright starcounts are not
very sensitive to the adopted value of the differential reddening
beyond the last node point, for the reasons explained in Sect.~3.2.

A change of the adopted Galactocentric distance within considerable
limits does not have an impact on the model counts, for reasons that
have been explained in Sect.~3.3. Also, the effect of scale-heights is
either minimal, or yields a systematic trend regardless of the region
being modeled, and therefore a change in this parameter is not
warranted by the present analysis (see below though). We therefore
infer that the current model implementation is not inconsistent with
the observed counts.

The distance of the Sun from the Galactic plane, and the scale-length
of disk stars have, both, a large impact on the predicted counts. The
question is whether a better {\it representation} of the observed
counts is provided by a suitable choice of $Z_{\sun}$ and $H_{\rm
Disk}$, or by a large scale-height of the reddening material. We
explore this by assuming $h_{\rm red}=110$~pc and by solving,
simultaneously, for $Z_{\sun}$ and $H_{\rm Disk}$. This simultaneous
solution is only possible because, as it can be seen from Figs.~16
and~17, Regions~2, 4, and~5 are mostly sensitive to $Z_{\sun}$ (but
not to $H_{\rm Disk}$), while Regions~1 and~3 are sensitive to
both. We use an iterative procedure in which we solve for $Z_{\sun}$
from Regions~2, and~4, apply this solution to Region~1 from which we
determine $H_{\rm Disk}$, which is used in turn to compute counts in
Regions~2, and~4, and so on until convergence. At each iteration, the
relevant parameters are computed from a full $\chi^2$ analysis of the
observed and computed starcounts in the magnitude range $8 \le V <
V_{\rm lim}$. Region~5 was excluded when computing $Z_{\sun}$ because
it was found to give a somewhat discrepant value for this parameter
with respect to the other two regions, as well as a larger error. For
example, from Runs~3 and~10 we get $Z_{\sun}=24 \pm 11 \,(3
\sigma)$~pc from Region~2, $Z_{\sun}=30 \pm 8 \,(3 \sigma)$~pc from
Region~4, and $Z_{\sun}=44 \pm 19 \,(3 \sigma)$~pc from
Region~5. Indeed, the mean from Regions~2 and~4 is not incompatible
(at the the $3 \sigma$ level) with the results from Region~5, given
its larger uncertainty. As for the disk's scale-height, we could not
use Region~3 because it never converged. Indeed, as it can be seen
from Figs.~16 and~17, whatever combination of $Z_{\sun}$ and $H_{\rm
Disk}$ underestimates the counts, and this is true even for the run
using Arenou's et al. (1992) model, as can be seen from Fig.~15.

The resulting parameters are $Z_{\sun}= 27 \pm 3$~pc and $H_{\rm Disk}
= 6.0 \pm 2.0$~kpc, where the uncertainties are at the 99\% confidence
interval (or, equivalently, $3\sigma$), and have been computed from
the $\chi^2$ fits using the parameter estimation scheme described by
Lampton et al. (1976). It must be emphasized that these results were
computed using only Regions~1,~2, and~4, for the reasons explained
above.

With the above determined values for $Z_{\sun}$ and $H_{\rm Disk}$,
the predicted counts indicate an overall uncertainty for {\it all}
regions (excluding, as always, Region~6, but including Regions 3 and 5
which were not used in the fit) of only 8\% in the counts (see
Table~5). This run is illustrated in Fig.~18. When comparing this run
with Run~8 (Table~4) we see that we have indeed reduced the scatter
from about 15\% to half this value with this choice of values for
$Z_{\sun}$ and $H_{\rm Disk}$.

Our long scale-height is compatible with the results by van der Kruit
(1986), who obtained $H_{\rm Disk} = 5.5 \pm 1.0$~kpc from his
analysis of the {\it Pioneer~10} optical integrated starlight
measurements. Ruphy et al. (1996), have suggested that van der Kruit's
result is sensitive to the adopted scale-height for disk
stars. However, if we adopt our lower set for the scale-height of disk
stars, then the Solar height above the plane is affected quite
considerably, but the scale-length remains unchanged. In fact, in this
case, we derive $Z_{\sun}= 8.2\pm 1.0$~pc and $H_{\rm Disk} = 6.5 \pm
1.8$~kpc, in disagreement with Ruphy's et al. claim that a smaller
scale-height will substantially reduce the derived scale-length. We
must emphasize that Ruphy's et al. IR counts, van der Kruit's optical
integrated light measurements, and our GSC counts are all mostly
sensitive to G and K giants, so that we are sampling basically the
{\it same} population of disk objects.

\begin{figure}
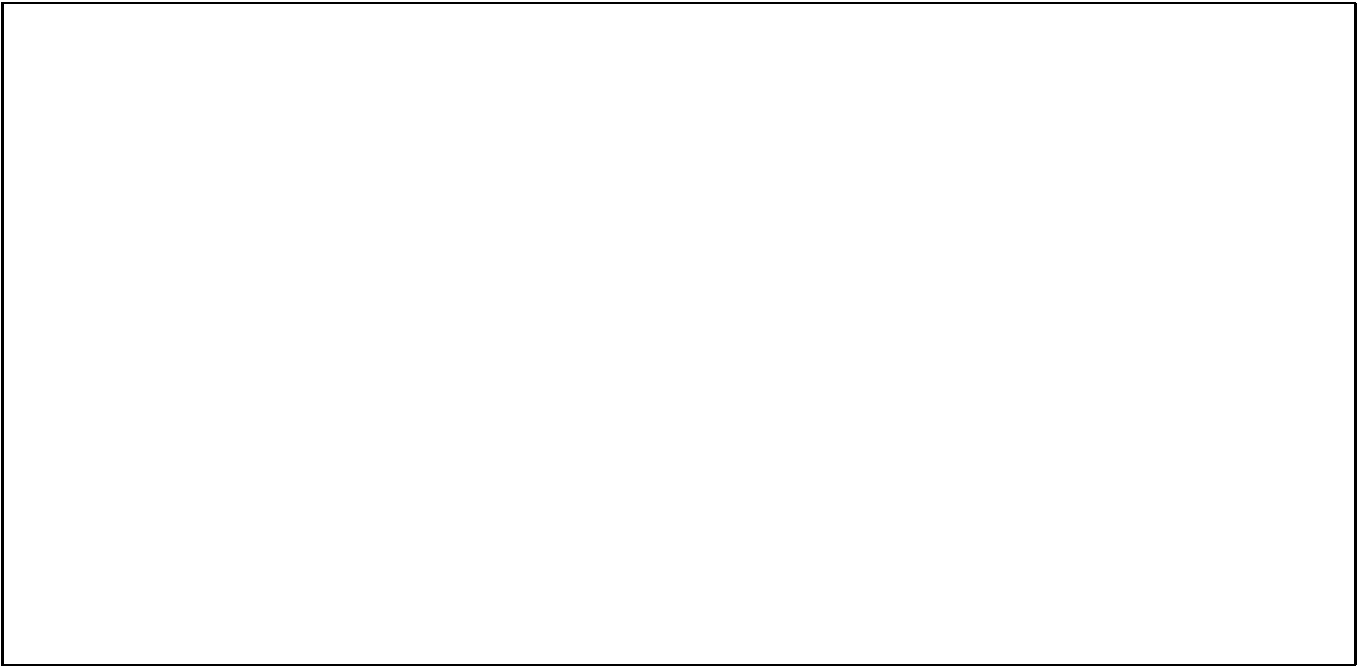

\picplace{8.8 cm}
\caption[]{Same as Fig. 13, except that we have used our best-fit
solution with $Z_{\sun}=+27$~pc and $H_{\rm Disk}=6$~kpc (solid
line). For comparison we also indicate Run~8 as a dashed line.}
\end{figure}

It is evident that more constraints on the models and a better insight
on stellar populations {\it in the disk} would be obtained if we had
colors. Unfortunately, the availability of colors from the
digitization of the short-exposure red plates obtained with the
UK-Schmidt at low-Galactic latitudes is not foreseen in any of the
current major digitization efforts (van Altena et al. 1993). This
situation is particularly unfortunate because of the need for an
optical counterpart to the near-infrared surveys of the Galactic plane
such as DENIS (Epchtein 1994), 2MASS (Skrutskie et al. 1995), and the
Two Micron Galactic Survey (Garzon et al. 1994).

\section{Conclusions}

We have presented the first systematic study of low-Galactic latitude
starcounts from the perspective of a Galactic structure and reddening
model, and we have demonstrated that our reddening model can be used
to obtain meaningful starcount estimates in the plane of the
Milky-Way. We find that it is possible to predict starcounts in the
range $8.0 \le V< 13.7$ with a mean accuracy of 15\% or better for
total samples of several thousand stars at $|b| < 10^{\degr}$, and for
total reddening at 1~kpc of up to $E(B-V) \sim 0.7$~mag. This accuracy
is similar to the accuracy with which starcounts at high galactic
latitudes can be modeled due to fluctuations in the stellar density
along the line-of-sight (Bahcall 1986). This opens the possibility of
using our reddening and starcount model to estimate and compare
observations of diffuse starlight to constrain the major structural
parameters of the Galaxy, as described by van der Kruit (1986,
1990). This is also an important step in determining the contribution
of diffuse Galactic light to the extragalactic diffuse light component
whose distribution is important as a test of Cosmological models
(Paresce 1985), for theories of galaxy formation and evolution (Cole
et al. 1992, Sasseen et al. 1995, Vaeisaenen 1996), and for theories
of growth of structure in the Universe (Bowyer \& Leinert 1990). In a
subsequent paper we will explore this issue in detail.

Another important application in the works is the use of our model to
compute the distance distribution of stars self-consistently with the
starcounts, in order to derive corrections to absolute parallax (van
Altena 1974, 1986, van Altena et al. 1988) which might in turn yield
better parallaxes than those contained in the latest edition of the
Yale Parallax Catalogue (van Altena et al. 1995).

By performing a simultaneous fit to three selected regions, we derive
a distance of the Sun from the symmetry plane of the Galaxy of $27 \pm
3 \, (3 \sigma)$~pc, while the disk's scale-length is found to be $6
\pm 2 \, (3\sigma)$~kpc.  The derived value for $Z_{\sun}$ is
sensitive to the adopted scale-height of disk stars, but no big
dependency of $H_{\rm Disk}$ on this parameter is found. These values
should be taken, of course, with caution because of the inherent
uncertainties in modeling optical starcounts in the presence of patchy
reddening material, which introduce the uncertainties explored in this
paper.

We have only grasped the rich potential for Galactic structure studies
that could be carried out with the digitized supplemental survey plate
collection of the GSC. It is clear from the analysis presented in this
paper that the problem is, by nature, multi-variate, and that a number
of parameters must be adjusted to a set of observables
simultaneously. The model evaluations described above are
computationally very intensive, and we are working on ways to make the
problem more tractable so that more regions can be compared to the
model predictions at once. A related issue is, of course, the choice
of optimum algorithms to arrive at the best overall solution. We are
confident that this is doable, since important advances have been made
in these two areas recently, as shown by Larsen (1996), Larsen et
al. (1996), and Chen (1996, 1997).

The reddening model source code (written in f77) may be obtained upon
request from the authors.

\begin{acknowledgements}
The four data files and the source code to compute the total reddening
from the Burstein \& Heiles (1982) maps at any requested point in the
sky were kindly made available to us by Dr. David Burstein. The
authors would like to acknowledge a number of constructive comments
from the referee, Dr. Helge J{\o}nch-S{\o}rensen, which lead to the
introduction of the more self-consistent Model-B for reddening
extrapolation, and several other suggestions from the referee which
have made the paper more readable overall. R.A.M. has greatly
benefited from conversations with Dr. Fernando
Comeron. R.A.M. acknowledges a grant from Allied Signal Aerospace
Guidance and Control Systems for the development of the reddening
model in the context of the Hubble Space Telescope Fixed-Head Star
Trackers and, in particular, to Arthur J. Bradley for his interest and
constant support during this research. R.A.M. also acknowledges a
travel grant (C-51073) from the Chilean Andes
Foundation. W.F.v.A. acknowledges support from NSF, NASA and the US
Naval Observatory, Flagstaff Station, the later for supporting the
development of the initial low-latitude reddening model in 1987 and
1988. The Guide Star Catalog was produced at the Space Telescope
Science Institute under U.S. Government grant. These data are based on
photographic data obtained using the Oschin Schmidt Telescope on
Palomar Mountain and the UK Schmidt Telescope. The Oschin Schmidt
Telescope is operated by the California Institute of Technology and
Palomar Observatory. The UK Schmidt Telescope was operated by the
Royal Observatory Edinburgh, with funding from the UK Science and
Engineering Research Council (later the UK Particle Physics and
Astronomy Research Council), until 1988 June, and thereafter by the
Anglo-Australian Observatory. The blue plates of the southern Sky
Atlas and its Equatorial Extension (together known as the SERC-J), as
well as the Equatorial Red (ER) were all taken with the UK Schmidt.
\end{acknowledgements}

\end{document}